\documentclass[10pt,oneside,onecolumn,aps,pra,preprintnumbers,bibnotes10pt,superscriptaddress,amsmath,amssymb]{revtex4-2}

\usepackage{graphicx}
\usepackage{float}
\usepackage{braket}
\usepackage{bbold}
\usepackage{amsmath}
\usepackage{amsthm}
\usepackage[dvipsnames]{xcolor}
\usepackage{multirow}
\usepackage{tikz}
\usetikzlibrary{positioning}
\usetikzlibrary{decorations.pathmorphing, decorations.pathreplacing}
\usepackage{ragged2e}
\usepackage{commath}
\usepackage{comment}
\usepackage{adjustbox}
\usepackage{subfig}
\usepackage{hyperref}
\usepackage[normalem]{ulem}
\usepackage{figchild}

\usepackage{epstopdf}
\usepackage{epsfig}
\usepackage{xcolor}
\usepackage{amsfonts}
\usepackage{mathptmx}
\usepackage{bbold}
\usepackage{mathtools}
\usepackage{bbm}
\usepackage{dsfont}

\theoremstyle{plain}

\theoremstyle{definition}

\theoremstyle{remark}

\newcommand{\idop}{\mathbbm 1}
\newcommand{\average}[1]{\langle #1 \rangle}
\newcommand{\Trace}{{\rm Tr}}

\newcommand{\re}{{\rm Re}}

\newcommand{\work}{\mathcal{W}}
\newcommand{\Deltau}{\Delta \mathcal{U}}

\def \ket#1{\mathinner{|{#1}\rangle}}

\newcommand{\ketbra}[2]{{\mathinner{| {#1} \rangle \langle {#2} |}} }

\newcommand{\matrixel}[3]{{\mathinner{\langle{#1}| {#2} | {#3}\rangle}} }

\newcommand{\G}{\mathcal{G}_\lambda}
\newcommand{\Prob}{\mathcal{P}}

\newcommand{\MU}{\hat{U}}
\newcommand{\hatH}{\hat{H}}
\newcommand{\KND}{K}
\newcommand{\Time}{\mathcal{T}}

\begin{document}

\title{Quantumness certification via non-demolition measurements}

\author{Paolo Solinas}
\affiliation{Dipartimento di Fisica, Universit\`a di Genova, and INFN - Sezione di Genova, via Dodecaneso 33, I-16146, Genova, Italy.}

\author{Stefano Gherardini}
\affiliation{Istituto Nazionale di Ottica del Consiglio Nazionale delle Ricerche (CNR-INO), Largo Enrico Fermi 6, I-50125 Firenze, Italy.}
\affiliation{European Laboratory for Non-linear Spectroscopy, Universit\'a di Firenze, I-50019 Sesto Fiorentino, Italy.}

\begin{abstract}
The fundamental question of when a static or dynamic system should be deemed intrinsically quantum remains a challenge to address in absolute terms. In this regard, a critical requirement lies in the certification (ideally, in real-time) of the emergence and persistence of genuine quantum features, principally entanglement and quantum superposition. Quantum Non-Demolition Measurements (QNDM) serve as the appropriate instrument for this certification, both from a theoretical and experimental standpoint. In this review paper, we explain, with accessible clarity, how the implementation of QNDM can be directly linked to a necessary and sufficient condition for the presence of genuinely quantum features in the system's state monitored over time in finite-dimensional systems, establishing a conceptual parallel with Leggett-Garg inequalities. Using concrete examples that detail the detection of negative terms in the quasi-probability density function resulting from QNDM, we introduce the core concepts for quantumness certification. As specific examples, we discuss an application where the quantum-to-classical transition due to the interaction with an environment can be tracked by QNDM. Moreover, we argue about the robustness of QNDM protocols in the presence of noise sources and their advantages with respect to standard Leggett-Garg inequalities defined by two-time correlators.
\end{abstract}

\maketitle

\tableofcontents

\section{Introduction}
\label{sec:Intro}

Classical and quantum systems are intrinsically different in their properties and behaviours. Despite these evident differences, discriminating the quantum features of a system is extremely challenging because they are elusive and highly fragile, especially when put in contact with an external environment. Indeed, both the interaction with the environment and/or the measurement apparatus to extract information perturb the system and generally lead to the destruction of these features. Nevertheless, determining whether a system behaves in a classical or a quantum way is of paramount importance, both for the foundations of quantum mechanics and for the development of quantum technologies.

The main quantum features that cannot be developed by any classical system are entanglement and superposition (or interference). They are related to the concepts of nonlocality and macrorealism (MR), respectively. Traditionally, the criteria used to distinguish between classical and quantum systems, thus identifying the presence of entanglement and quantum superposition, are the famous Bell's inequalities~\cite{Bell1964,Freedman1972,Aspect1982,Pan1998} and the Leggett-Garg inequalities (LGIs)~\cite{Leggett1985}.  
The former focuses on nonlocality and quantum correlations related to entanglement, while the latter concerns quantum superposition.

Macrorealism is a concept introduced by Leggett and Garg in 1985~\cite{Leggett1985}, whereby a classical system must satisfy three main assumptions:  
$1)$ {\it Macrorealism per se} (MRps), which states that the system is at all times in one of the states available to it;  
$2)$ {\it Noninvasive Measurability} (NIM), which requires that it should be possible to determine the state of a system without disturbing its state and thus its subsequent dynamics; and  
$3)$ {\it Induction}, added and clarified later~\cite{Halliwell2016}, which asserts that future measurements cannot influence the present state.

By measuring time correlations between sequential measurements at two times, Leggett and Garg derived the set of inequalities~\cite{Leggett1985} known as LGIs, which are formally similar to Bell’s inequalities. If these inequalities are not satisfied, then the system violates the MR and therefore exhibits quantum features. LGIs, like Bell’s inequalities, provide only a sufficient condition for the violation of MR. Indeed, there are situations in which a system violates MR, i.e., it has quantum properties, while still satisfying the LGIs. The conjunction of the MR's assumptions 1)-3) listed above can be tested experimentally, in strict relation to observation of LGIs' violations (see, for example Refs.~\cite{DresselPRL2011,Knee2012,Knee2016,ChatterjeePRL2025}).

In this review, we examine some recent studies that have been developed to address this issue by introducing an experimentally feasible and powerful protocol known as Quantum Non-Demolition Measurements (QNDM). This approach builds on measurement schemes proposed at the end of 1970s~\cite{ThornePRL1978,Braginsky1980,BraginskyBook,Caves1980,CavesRevModPhys}, which have been modified and adapted to extract information from sequential non-projective measurements~\cite{Onofrio1995,Calarco1995,Calarco1997,solinas2013work,solinas2015fulldistribution,solinas2016probing,solinas2021,solinas2022,DeChiara2018,GherardiniTutorial,solinas2024}.

The key idea of QNDM is to store the information about a desired observable in the phase of a quantum detector that is coupled to the system. The coupling is designed so that it leaves the statistics of these observables unchanged. Then, after a sequence of system-detector interactions, the phase of the detector is measured, and from it one can reconstruct a quasi-probability density function of the measurement outcomes. The latter is not always positive, and it can be proven~\cite{solinas2024} that the presence of negative regions is unequivocally associated with the presence of genuinely quantum features in the system's state. More precisely, the quasi-probability density function has negative regions if and only if MRps is violated, namely, at least a quantum superposition of the observable's eigenstates is identified in the system's state at the time instants when system-detector interaction occurs. On the other hand, if we find that the quasi-probability density function resulting from QNDM is positive, then we can show that such a distribution can be obtained from sequential projective measurements, and thus described by the so-called Two-Point Measurement (TPM) protocol~\cite{esposito2009nonequilibrium,esposito2009Erratum,campisi2009fluctuation, campisi2011erratum}. It is known that the results provided by the TPM scheme have a classical-quantum correspondence~\cite{allahverdyan2014,GherardiniTutorial,solinas2024}.

Similar ideas of using an additional quantum system to reconstruct features of a quantum system have also been proposed in the field of full-counting statistics~\cite{levitov1993charge,levitov1996electron,nazarov2003full}, where the statistics of electrons flowing through a wire is measured. These results led to several implementations in condensed matter and electronic circuits~\cite{Reulet2003,Timofeev2007,Maisi2014}, bringing advancement in such fields. While it was known that the approach of full-counting statistics entails a quasi-probability distribution~\cite{nazarov2003full, clerk2011full, bednorz2010quasiprobabilistic, bednorz2012nonclassical,Belzig2001}, implications in terms of macrorealism violation and the connection of the latter to the negativity of the quasi-probability distribution have not been fully understood.

The article is organized as follows.  
In Sec.~\ref{sec:Work_statistics}, we present a physical scenario in which we aim to extract information about work and internal energy variation of a quantum system.  
This is a prototypical example where the QNDM approach gives a consistent advantage over the standard projective measurements.
It also provides the basis for Sec.~\ref{sec:QNDM_2measurement}, where we introduce the QNDM protocol that is designed to preserve interference effects and quantum superposition. 
In Sec.~\ref{sec:macrorealism_and_LGI}, we discuss the main properties of the QNDM protocol and its relation with the LGIs, and we quantify the resources needed for their implementation (Sec.~\ref{sec:QNDM_robustness}). Then, we report an example where QNDM can be used to track the quantum-to-classical transition induced by an external environment (Sec.~\ref{sec:quantum_to_classical}), showing that the protocol is both robust against noise and efficient in terms of resources. Finally, Sec.~\ref{sec:Conclusions} contains the conclusions and outlines future perspectives.  

\section{Motivation: Describing internal energy variation and work at the quantum level}
\label{sec:Work_statistics}

The introduction of QNDM in recent literature~\cite{solinas2015fulldistribution, solinas2016probing, solinasPRA2017,solinas2021, solinas2024} comes from a practical need: to give a clear and operational definition of physically relevant quantities such as work, heat, and internal energy variation at the quantum level.

Work, heat, and internal energy variation are cornerstone concepts in classical thermodynamics and mechanics. It is thus natural to ask how one can define or measure them when the dimension of the system under scrutiny decreases, and quantum effects become relevant. Let us focus on the mechanical definition of internal energy variation and work by considering a system that is isolated from the environment (i.e., it does not dissipate energy in terms of heat) and is driven by an external force that does work on the system. In such a case, the work $\work$ done by the external force is equal to the variation of the internal energy $\Deltau$: $\work = \Deltau$.

However, when we reach the quantum regime, we face a problem whenever we need to measure and characterize the statistics of physical quantities defined at two or multiple times. In classical physics, we implicitly assume that we are able to measure these physical quantities without perturbing the system at any time. Instead, in quantum mechanics, the measurement procedure necessarily perturbs the system by inducing a collapse of its wave function~\cite{GherardiniTutorial}.

From a classical perspective, we can measure sequentially the energy of the system at the beginning $\mathcal{E}_i$ and at the end $\mathcal{E}_f$ of the driving protocol, and taking their difference: $\work = \Deltau = \mathcal{E}_f - \mathcal{E}_i$. All the physical quantities are well defined, and we thus have a clear theoretical definition that matches the measuring procedure.

In the quantum case, we suppose to know the time-dependent Hamiltonian $\hatH(t)$ that defines the driving protocol. The time dependence of $\hatH(t)$ is due to the presence of a classical external time-dependent force that exchanges energy with the quantum system, by doing work on it. The system evolves under $\hatH(t)$ from time $t=0$ to time $t=\Time$, undergoing a unitary transformation, which we denote as $\hat{U}(\Time) \equiv \hat{U}$. The initial state of the system is supposed to be described by the density operator $\hat{\rho}^0$. As in the classical case, if the system is isolated, then the work is equal to the variation of the internal energy, but with a caveat. Indeed, the request is that the initial and final energies of the classical system are substituted by the statistical averages (expectation value) obtained from measuring the quantum system at times $t=0$ and $t=\Time$:
\begin{equation}
    \work = \Deltau = \average{\hat{H}(\Time)} - \average{\hat{H}(0)} = \Trace\Big[ 
    \left( \hat{U}^\dagger \hat{H}(\Time) \hat{U} - \hat{H}(0) \right) \hat{\rho}^0 \Big].
    \label{eq:W_def}
\end{equation}

Looking at Eq.~\eqref{eq:W_def}, one could define the work operator $\hat{W} := \hat{U}^\dagger \hat{H}(\Time) \hat{U} - \hat{H}(0)$~\cite{allahverdyan2014,Deffner2016,CerisolaNatComm2017} that is Hermitian and allows us to interpret the average work as its trace over the initial density operator $\hat{\rho}^0$.
However, the use of such an operator requires careful consideration as it hides a fundamental issue. Indeed, the average $\work$ is related to two measurements performed sequentially at different times. This means that, even if $\hat{W}$ is a Hermitian operator, it cannot be associated in general with a sequential measurement process since the latter is always local in time, at least in the usual projective measurement assumption~\cite{roncaglia2014work}.

Initially, this issue has been circumvented because the first studies on a driven quantum system considered initializing the system in a thermal state built over the initial Hamiltonian $\hat{H}(0)$, for which sequential measurements suffice to determine two-time correlators. In such cases, one can adopt the TPM protocol~\cite{talkner2007fluctuation, engel2007jarzynski,esposito2009nonequilibrium, esposito2009Erratum,campisi2009fluctuation, campisi2011erratum,KafriPRA2012,HernandezGomezPRR2020,Gherardini2022energy,PhysRevE.111.014139}. This consists of measuring the system energy at the initial time $t=0$, letting the system evolve under the external drive, and then measuring the energy at time $t = \Time$. If we denote the eigenvalues and eigenvectors of $\hat{H}(t)$ with $\epsilon_j(t)$ and $\ket{\epsilon_j(t)}$, respectively, it can be shown~\cite{talkner2007fluctuation,engel2007jarzynski} that the statistics of the work (or internal energy variation) is given by the probability distribution
\begin{equation}
    \Prob(W) = \sum_{i,j} \delta\Big[ W - \left( \epsilon_j(\Time) - \epsilon_i(0) \right) \Big] P_i^{(0)} P^{(\Time)}_{i \rightarrow j}\,, 
    \label{eq:TMP_distribution}
\end{equation}
where $P_i^{(0)} := \langle\epsilon_i(0)|\hat{\rho}^0|\epsilon_i(0)\rangle$ is the probability to find the system initially in the state $\ket{\epsilon_i(0)}$, and $P^{(\Time)}_{i \rightarrow j} := |\langle\epsilon_j(\Time)|\hat{U}|\epsilon_i(0)\rangle|^2$ is the probability for the occurrence of the transition from $\ket{\epsilon_j(0)}$ to $\ket{\epsilon_j(\Time)}$ due to the unitary transformation $\hat{U}$. The average work provided by the TPM protocol can be obtained by the distribution \eqref{eq:TMP_distribution} and reads as
\begin{equation}
    \work = \sum_{i,j} \left[ \epsilon_j(\Time) - \epsilon_i(0) \right ] P_i^{(0)} P^{(\Time)}_{i \rightarrow j}.
    \label{eq:TMP_average}
\end{equation}
Eqs.~\eqref{eq:TMP_distribution} and \eqref{eq:TMP_average} have an immediate interpretation in simple statistical terms.
The work done $\epsilon_j(\Time) - \epsilon_i(0)$ during the process is weighted by the probability to start from the state $\ket{\epsilon_j(0)}$ and to end in the state $\ket{\epsilon_j(\Time)}$, exactly as it happens in a classical statistical process.

Although its direct application, the first measurement of the TPM protocol causes a perturbation on the system, which leads to the collapse of the initial (at time $t=0$) system's wave-function. This ultimately entails a change in the dynamics and in the expected prediction~\cite{GherardiniTutorial}. 
To clarify this point, we consider the definition of the average work as given by Eq.~\eqref{eq:W_def} and show that, for a {\it generic non-diagonal} initial density operator $\hat{\rho}^0$, the TPM protocol cannot recover the full (unperturbed) expression of the average work. Indeed, the right-hand side of Eq.~\eqref{eq:W_def} admits two contributions~\cite{solinas2013work}: one related to the diagonal elements of $\hat{\rho}^0$, while the other depends on the corresponding off-diagonal elements. Formally, we have:
\begin{equation}
    \work = \Deltau = \sum_{i,j} \left[ \epsilon_j(\Time) - \epsilon_i(0) \right ] P_i^{(0)} P^{(\Time)}_{i \rightarrow j} +
    \sum_{i\neq k,j} \epsilon_{j}(\Time) \rho^0_{i,k}  U^\dagger_{k,j} U_{j,i} \,,
    \label{eq:W_interference}
\end{equation}
where $\sum_{j}P^{(\Time)}_{i \rightarrow j}=1$ by construction, $\rho^0_{i,k} := \matrixel{\epsilon_{i}(0)}{\hat{\rho}^0}{\epsilon_{k}(0)}$, $U_{j,i} := \matrixel{\epsilon_{j}(\Time)}{ \hat{U} }{\epsilon_{i}(0)}$, $U^\dagger_{k,j} := \matrixel{\epsilon_{k}(0)}{\hat{U}^\dagger}{\epsilon_{j}(\Time)}$. From this calculation, we evince that the difference between the unperturbed average work \eqref{eq:W_interference} and the average work returned by the TPM protocol \eqref{eq:TMP_average} is in the contributions associated with the off-diagonal terms $\rho^0_{i,k}$ of the initial density operator~\cite{solinas2013work}. These contributions stem from the interference or coherent superposition of the eigenstates of the initial Hamiltonian $\average{\hat{H}(0)}$~\cite{solinas2013work,solinas2015fulldistribution, solinas2016probing}.

This observation allows us to make a parallel with the famous two-slit experiment with quantum particles ~\cite{feynman1965quantum}. If both slits are open, we observe an interference pattern of the particle arriving at the screen. If one of the two slits is closed or we perform a measurement to establish through which slit the particle has passed, the interference pattern vanishes. In the present discussion, the initial projective measurement of $\average{\hat{H}(0)}$ leads to the collapse of the system's wave-function at $t=0$ and, as a consequence, forces the system into one of the eigenstates of $\average{\hat{H}(0)}$. As in the two-slit experiment, the first measurement of the TPM protocol destroys the interference patterns and changes the subsequent dynamics of the system.

The challenge of measuring the variation of internal energy in a quantum system without destroying the quantum coherence and/or correlations, initially present in the system's state or generated during its dynamics, has been a subject of recent discussions in the literature. Specifically, pushed by no-go theorems~\cite{Perarnau-Llobet2017No-Go,LostaglioKirkwood2022} that identify fundamental constraints on measuring physical quantities at two consecutive times, several approaches have been introduced. These include non-linear schemes using information about the initial system's state~\cite{MicadeiPRL2020,MicadeiPRL2021,GherardiniPRA2021,HernandezGomez_entropy_coherence,GiananiQST2023}, inference protocols with initial-time measurement~\cite{Deffner2016,SonePRL2020}, procedures implementing weak measurements~\cite{bednorz2010quasiprobabilistic,BizzarriQST2025}, and quasi-probability approaches~\cite{hofer2016,Lostaglio2018,Potts2019,Levy2020,solinas2021,solinas2022,LostaglioKirkwood2022,hernandez2022experimental,HernandezNpjQI2024,ArvidssonShukur2024review,PRXQuantum.5.030355,PezzuttoQST2025}. Among the latter, we count Quantum Non-Demolition Measurements (QNDM)~\cite{Braginsky1980,BraginskyBook,Caves1980, CavesRevModPhys} that have been specifically designed to acquire information about a quantum system at multiple times while preserving the full quantum dynamics of the system.

\section{Quantum non-demolition measurement}
\label{sec:QNDM_2measurement}

To introduce the QNDM scheme, it is preferable to step back from a specific application and consider a more general framework~\cite{solinas2024}.
Subsequently, we will show how this framework applies to the measurement of the internal energy variation.

Let us thus suppose we have a quantum system evolving under a unitary transformation in a time interval $t_0 \leq t \leq \Time$. At times $t_0\leq t_1 = \Time$, we measure a generic observable $\hat{A}$ (Hermitian operator). In order to realize QNDM, we consider an auxiliary quantum detector to store the desired information in the phase of the detector's state, which is eventually measured. This scheme allows us to preserve the quantum coherence in the initial density operator of the system, and derive the quantum correlation function of observables measured at multiple times, without the result being perturbed by the interaction with the detector~\cite{solinas2015fulldistribution, solinas2016probing,solinasPRA2017,solinas2021,solinas2022}.

Ideally, the sequential non-demolition measurements assume fast system-detector couplings so that, during the interaction with the detector, the system dynamics can be considered frozen. The main point in the QNDM approach is to choose a proper system-detector coupling. If we want to measure the observable $\hat{A}$ we need to implement the unitary transformations $\hat{u}_k = \exp\{ i (\lambda_k/2) \hat{A} \otimes \hat{p} \}$ at any time $t_k$ with $k=0,1$~\cite{solinas2013work,solinas2015fulldistribution,solinas2016probing}. Here, $\lambda$ is an effective coupling, and $\hat{p}$ is an operator acting on the detector's degrees of freedom. The unitary transformations $\hat{u}_k$ are taken in order to not induce any transition in the system state with respect to the basis that spectrally decomposes $\hat{A}$, whose statistics are thus preserved. Hence, the effect of the system-detector coupling is to make a phase accumulate in the detector's state between the eigenstates $\hat{p}$ of $\hat{A}$.

Between the couplings at times $t_k$, the system evolves freely from the detector. This means that there is no restriction on the system's time evolution that can also be driven by an external time-dependent field, which performs work on the system as in Sec.~\ref{sec:Work_statistics}.
Such an evolution from time $t_{0}$ to $t_1$ is described by the unitary operator $\MU_1 \otimes \idop = \MU( t_{1}, t_{0} ) \otimes \idop$ so that the total unitary transformation acting on both the system and detector is $\MU_{tot} = \hat{u}_1 (\MU_1 \otimes \idop) \hat{u}_0$. To keep the notation as simple as possible, we  denote $ \MU_i \equiv \MU_i \otimes \idop$ so that $\MU_{tot}$ reads as 
\begin{equation}
\MU_{tot} =  e^{ i \frac{\lambda_1}{2} \hat{A} \otimes \hat{p} } \MU_1 \, e^{ i \frac{\lambda_0}{2} \hat{A} \otimes \hat{p} }.
\label{eq:U_tot}
\end{equation}
This theoretical framework also accounts for sequential measurements of different operators.
Indeed, considering that $[\hat{U}_1,\hat{A}] \neq 0$ in the general case, $\MU_{tot}$ can be equivalently expressed as 
\begin{equation}
\MU_{tot} = \MU_1 e^{ i \frac{\lambda_1}{2} \hat{B} \otimes \hat{p} } \, e^{ i \frac{\lambda_0}{2} \hat{A} \otimes \hat{p} },
\end{equation}
where $\hat{B} := \MU_1^\dagger \hat{A} \MU_1$.

To understand how the QNDM approach works, it is convenient to decompose the initial state of the system in the basis $\{ \ket{i} \}$ in which the observable $\hat{A}$ is diagonal, i.e., $\hat{A} \ket{i} = a_i \ket{i}$ with $a_i$ real numbers, so that $\ket{\psi_0} = \sum_i  \psi^0_i \ket{i}$. 
At the same time, we must initialize the detector's state in a coherent superposition of eigenstates of $\hat{p}$, so that it is taken equal to $\ket{\phi_0} = \left(1/\sqrt{M} \right) \sum_p \ket{p}$. This assures that, during the system-detector coupling, a phase is accumulated in the state of the detector between the eigenstates $\ket{p}$. Given the initial state $\ket{\psi_0}\ket{\phi_0} $ and the unitary transformation $\MU_{tot}$, the final system-detector's state is (see Ref.~\cite{solinas2016probing,DeChiara2018,solinas2024} for analogous calculation)
\begin{equation}\label{eq:QNDM_dynamics}
     \ket{\Psi} = \frac{1}{\sqrt{M}}\sum_p \sum_{i, j} e^{i \frac{\lambda p}{2}( a_j + a_i)} U_{1,ji} \psi^0_{i} \ket{j} \ket{p},     
\end{equation}
where here $U_{1,ji} := \matrixel{j}{\MU_1}{i}$ and $\lambda_0=\lambda_1=\lambda$.

As the information on the evolution of the monitored system is extracted by measuring the phase accumulated in the detector's state (experimentally, it can be done using an interferometric technique~\cite{dorner2013extracting,mazzola2013measuring,HernandezNpjQI2024}), it is worth determining a formal expression for such a phase, whose meaning we are going to discuss. Denoting with $\hat{R} = \ketbra{\Psi}{\Psi}$ the total density matrix after the evolution, we trace out the system's degrees of freedom to obtain the final density matrix $\hat{r} = 
\Trace_S\left[ \hat{R} \right]$ of the detector.
Then, we take the accumulated phase 
\begin{equation}
\G = \frac{ \matrixel{p}{\hat{r}}{-p} }{ \matrixel{p}{\hat{r}^0}{-p} },
\end{equation}
where $\hat{r}^0 := \ketbra{\phi_0}{\phi_0}$.

The function $\G$ defines the QNDM {\it quasi-characteristic function} associated with the correlation function of $\hat{A}$ measured at times $t_0,t_1$. It can be shown, indeed, that the derivatives of $\G$ with respect to $\lambda$ at $\lambda=0$ are proportional to the moments of the distribution built with the eigenvalues of $\hat{A}$~\cite{solinas2015fulldistribution, solinas2016probing,solinasPRA2017,solinas2021,solinas2022}. Writing $\G$ as~\cite{solinas2013work,DeChiara2018,GherardiniTutorial}
\begin{equation}
    \G = \Trace \left[
    e^{ i \frac{\lambda}{2} \hat{A} } \MU_1 \, e^{ i \frac{\lambda}{2} \hat{A} } \hat{\rho}^0 
    e^{ i \frac{\lambda}{2} \hat{A} } \MU_1^\dagger \, e^{ i \frac{\lambda}{2} \hat{A}  }
    \right],
\end{equation}
taking the first derivatives of $\G$ with respect to $\lambda$ and setting $\lambda=0$, we obtain the expectation value of the variation of $\hat{A}$ within the time-interval $[t_0,t_1]$~\cite{solinas2013work,DeChiara2018}:
\begin{equation}
    (-i) \left . \frac{d \G}{d  \lambda}  \right |_{\lambda=0} = \Trace \Big[ \left( \hat{\rho}(t_1) -  \hat{\rho}^0 \right) \hat{A} \Big] = \average{\hat{A}(t_1)} - \average{\hat{A}(t_0)} = \average{ \hat{A}^{H}(t_1) - \hat{A}(t_0) },
\end{equation}
where $\hat{\rho}(t_1) = \MU_1 \hat{\rho}^0 \MU_1^\dagger$, $\average{\hat{A}(t_1)} = \Trace [\hat{A} \hat{\rho}(t_1)]$ and $\hat{A}^{H}(t_1) := \MU_1^\dagger \hat{A} \MU_1$, with the superscript $H$ denoting the Heisenberg representation. Moreover, by direct calculation~\cite{DeChiara2018}, it can be shown that the second derivative of $\mathcal{G}_\lambda$ with respect to $\lambda$, computed at $\lambda=0$, is proportional to $\Trace\left[ \left( \hat{A}^{H}(t_1) - \hat{A}(t_0) \right)^2 \hat{\rho}^0 \right]$, which is the second moment of the variation of $\hat{A}$ at times $t_0,t_1$.\\
It should be stressed that these moments originated by the QNDM quasi-characteristic function $\G$ have a very specific time ordering. This is due to the fact that the operators $\hat{A}^{H}(t)$ at different times $t$ do not commute with each other. For example, in composing three operators using $\hat{A}(t_0)$ and $\hat{A}^{H}(t_1)$, we have combinations such as $\hat{A}^{H}(t_1)\hat{A}(t_0)^2$, $\hat{A}^{H}(t_1)^{2} \hat{A}(t_0)$, $\hat{A}(t_0)\hat{A}^{H}(t_1)\hat{A}(t_0)$ and so on. The specific measurement procedure entails the time ordering; in general, different measurement protocol gives rise to different characteristic functions $\G$ that differ in the time ordering entering the statistical moments of $\hat{A}^{H}(t_1) - \hat{A}(t_0)$.

With these clarifications and given the final state in Eq.~\eqref{eq:QNDM_dynamics}, the expression of the QNDM quasi-characteristic function $\G$ simplifies to
\begin{equation}\label{eq:characteristic_function}
    \G = \sum_{i, l, j} e^{i \lambda \left (a_j + \frac{a_i+a_l}{2} \right)}    U_{1,ji} \rho^0_{il} U^{*}_{1,lj}
\end{equation}
The right-hand side of Eq.~\eqref{eq:characteristic_function} can be directly obtained experimentally, as it is associated with protocols measuring a phase; in this sense, it has a clear operational meaning.

Starting from the observation that $\G$ is a quasi-characteristic function, it is natural to ask what kind of probability distribution can be associated with it. As usual, the corresponding probability distribution is obtained by taking the inverse Fourier transform of $\G$, i.e., $\Prob_{\rm QNDM}(\Delta) = (2 \pi)^{-1} \int \exp\{ -i \lambda \Delta \} \G d \lambda$. It turns out that, in general, $\Prob_{\rm QNDM}(\Delta)$ is not positively defined and, therefore, is a quasi-probability density function.
We must stress that $\Prob_{\rm QNDM}(\Delta)$ can be reconstructed experimentally; this means that the negative regions of $\Prob_{\rm QNDM}(\Delta)$ have a physical interpretation, as discussed below, albeit they are not associated with any physical observable.

To given an interpretation of the contributions in $\Prob_{\rm QNDM}(\Delta)$, let us introduce the function
\begin{equation}\label{eq:stochastic_Delta}
    \Delta_{j,i,l} = a_j + \frac{ a_i + a_l }{2}
\end{equation}
and the projectors $\hat{\Pi}_k = \ketbra{k}{k}$ that obey the relations $\hat{\Pi}_k^2 = \hat{\Pi}_k$, $\sum_k \hat{\Pi}_k = \idop$ and $\hat{\Pi}_k \hat{\Pi}_j = \delta_{kj}\hat{\Pi}_k$. With these notations, the QNDM quasi-probability density function can be formally written as
\begin{equation}
    \Prob_{\rm QNDM}(\Delta) = \sum_{i, l, j} P_{\rm QNDM}(j, i, l) \delta \left[ \Delta - \Delta_{j,i, l} \right],
	\label{eq:ND_distribution}
\end{equation}
where $\delta[\cdot]$ denotes the Dirac's delta and 
\begin{equation}
    P_{\rm QNDM}(j, i, l) =  U_{1,ji}  \rho^0_{il} U^{*}_{1,lj}  = \Trace \left[  \hat{\Pi}_j \MU_1 \hat{\Pi}_i \hat{\rho}^0 \hat{\Pi}_l \MU^\dagger_1 \right].
    \label{eq:ProbAmplitude}
\end{equation}
It is important to note that $P_{\rm QNDM}(j, i, l)$ contains the Kirkwood-Dirac one~\cite{GherardiniTutorial,ArvidssonShukur2024review}, which is obtained by summing over the index $l$ related to the quantum measurement with projectors $\{ \hat{\Pi}_l \}$.

\begin{figure}
    \begin{center}
    \includegraphics[scale=.6]{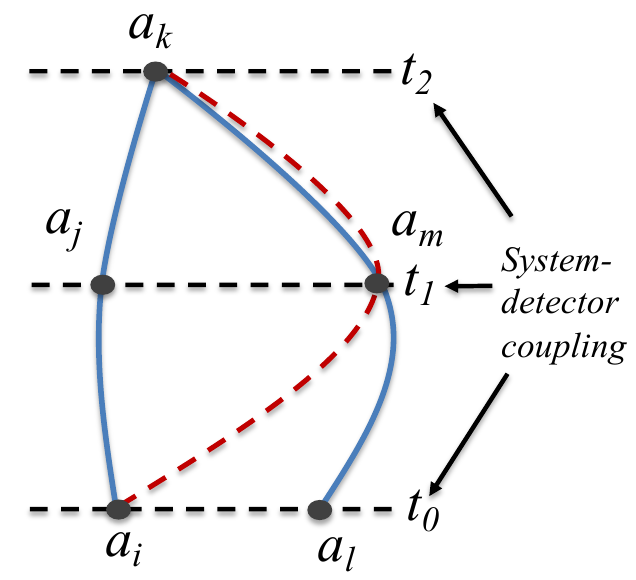}
    \end{center}
    \caption{Pictorial representation of the system's time evolution in terms of the outcomes from the sequential measurement at the three times $(t_0,t_1,t_2)$. In this picture, the trajectories developed by the measured system are ascribable to paths built over the measurement outcomes.
    The possible outcomes of the measurements are $a_i$ and $a_l$ at time $t_0$, $a_j$ and $a_m$ at time $t_1$, and $a_k$ at time $t_2$. The sequence of outcomes identifies the paths that the monitored system could follow during its time evolution.    
    The red dashed curve represents a classical path where, at any time, we have a single outcome (e.g., $a_i \rightarrow a_m \rightarrow a_k$) so that the observable has a determinate outcome. The blue curves present quantum paths in which the system is in a superposition of states with different outcomes. In such a case, the statistics of outcomes obtained from measuring $\hat{A}$ at times $(t_0,t_1,t_2)$ change as there is a coherent superposition of paths that violates the requirement of macrorealism per se.} 
    \label{fig:superpositions}
\end{figure}

In Eq.~\eqref{eq:ND_distribution}, the function $P_{\rm QNDM}(j,i,l)$ is the interference contribution between two alternative and superposed paths $i \rightarrow j$ and $l \rightarrow j$. A pictorial representation is presented in Fig.~\ref{fig:superpositions} where the measurement sequence is associated with a trajectory in the measurement outcomes space, e.g., $a_i \rightarrow a_j \rightarrow a_k$.

Since $P_{\rm QNDM}(j,i,l)$ are interference contributions, they may take negative values from which the negativity of $\Prob_{\rm QNDM}(\Delta)$ comes. The $P_{\rm QNDM}(j,i,l)$ can be even complex numbers with imaginary parts, but the latter do not contribute to the non-positivity of the QNDM quasi-probability density function $\Prob_{\rm QNDM}(\Delta)$ for symmetry reasons~\cite{GherardiniTutorial}.

To understand the implications of the quasi-probability density function \eqref{eq:ND_distribution}, let us separate its classical and quantum contributions.
The classical contributions $\Prob_{\rm cl}(\Delta)$ of $\Prob_{\rm QNDM}(\Delta)$ are the ones in which, during the whole evolution, the system is always in an eigenstate of the measured observable (see Fig.~\ref{fig:superpositions} for a pictorial representation); as a consequence, they satisfy the MRps condition presented in the Introduction (Sec.~\ref{sec:Intro}). These are the contributions that can be assessed through the TPM protocol that relies on sequential projective measurements.
The latter correspond to trajectories associated with a positive (multitime) probability, and thus ascribable to classical evolutions in statistical mechanics. Formally, they are obtained from $\Prob_{\rm QNDM}(\Delta)$ when $l=i$, i.e.,
\begin{equation}\label{eq:Pcl_2meas}
    \Prob_{\rm cl}(\Delta) = \sum_{i, j} P_{\rm QNDM}( j,  i, i) \delta \left[ \Delta - \Delta_{ j, i, i}   \right]
\end{equation} 
with
\begin{equation}\label{eq:PND_jii}
    P_{\rm QNDM}(j, i, i) = U_{1,ji}  \rho^0_{ii} U^{*}_{1,ij} = P^{(0)}_{i} P^{(1)}_{i \rightarrow j}\,.
\end{equation}
In Eq.~\eqref{eq:PND_jii}, $P^{(0)}_i = \rho^0_{ii}$ and $P^{(1)}_{i \rightarrow j} = |U_{1,ji}|^2$ are the probabilities of finding the system initially in the state $\ket{i}$ and then in the state $\ket{j}$ after the transition $\ket{i} \rightarrow \ket{j}$, respectively. This proves that $P_{\rm QNDM}(j, i, i)$ is a positive probability.

On the other hand, the quantum contributions $\Prob_{\rm q}(\Delta)$ of $\Prob_{\rm QNDM}(\Delta)$ are the ones for which the system is in a superposition of the eigenstates of $\hat{A}$ at any time $t$ of the whole evolution (see blue curves in Fig.~\ref{fig:superpositions}) and, in the terminology used in Sec.~\ref{sec:Intro} and Ref.~\cite{Leggett1985}, they are the ones that violate the MRps assumption. In the easiest picture with two measurement times (e.g., $t_0, t_1$), the quantum contributions originate from considering $l\neq i$, which reflects the fact that the system is initially in a coherent superposition of the eigenstates $\ket{i}$ and $\ket{l}$ of $\hat{A}$. Hence, formally,
\begin{equation}\label{eq:Pq_2meas}
     \Prob_{\rm q}(\Delta) = \sideset{}{'}\sum_{i, l, j} P_{\rm QNDM}( j, i, l) \delta \left[ \Delta - \Delta_{ j, i, l} \right],      
\end{equation}
where the notation $\sideset{}{'}\sum$ denotes the sum over the indices $i, l, j$ but neglecting the terms corresponding to $l = i$.

Let us now suppose that the measurement outcome $a_i$ can be only $\pm 1$. The classical contributions in Eq.~\eqref{eq:Pcl_2meas} are associated with $\Delta_{j, i, i} = a_j + a_i$ that assume the values $0$ and $\pm 2$.
On the contrary, the quantum contributions in Eq.~\eqref{eq:Pq_2meas} are associated with $\Delta_{j, i, l}$ with $l \neq i$ that also take the values $\pm 1$. Therefore, in this sense, the quantum contributions of $\Prob_{\rm QNDM}(\Delta)$ are linked with sequences of measurement outcomes that are forbidden in classical transitions. 

\subsection{Application to driven systems: Measurement of quantum work}
\label{sec:work_measurement}

The results in Sec.~\ref{sec:QNDM_2measurement} can be applied to discuss how the QNDM approach can be exploited to measure the work in a closed quantum system. To do this, we need to assume that at time $t=0$ the coupling between the system and the detector occurs with $\hat{A} = \hatH(0)$, and at time $\Time$ it is with $\hat{A} = \hatH(\Time)$. Moreover, in the second coupling transformation, we have to flip the coupling sign $\lambda \rightarrow - \lambda$, to store the variation of the system energy in the detector state during the composite evolution.

Following the results in Eqs.~\eqref{eq:ND_distribution}, \eqref{eq:Pcl_2meas} and \eqref{eq:Pq_2meas}, the quasi-probability density function of work reads as
\begin{eqnarray}\label{eq:QNDM_work_distribution}
    \Prob_{\rm QNDM}(W) = \sum_{i,j} \delta\Big[ W- \left( \epsilon_j(\Time) - \epsilon_i(0) \right) \Big] P_i^{(0)} P^{(\Time)}_{i \rightarrow j} + \sum_{i\neq l,j} \rho^0_{i,l} U_{l,j}^{\dagger} U_{j,i} \delta\left[ W - \Delta_{j,i,l} \right],    
\end{eqnarray}
where here $\Delta_{j,i,l} = \epsilon_j(\Time) - (\epsilon_i(0)+\epsilon_l(0))/2$.

The average work can be calculated from the derivative of the corresponding quasi-characteristic function with respect to $\lambda$, or directly from the distribution \eqref{eq:QNDM_work_distribution}. It turns out it is equal to the expected average in Eq.~\eqref{eq:W_interference}, which is not recovered by the TPM scheme whenever the initial density operator $\hat{\rho}^0$ is {\it not diagonal} along the eigenbasis of $\hat{H}(0)$~\cite{Perarnau-Llobet2017No-Go,LostaglioKirkwood2022}.
Importantly, we have observed that every time the TPM protocol fails to reproduce the unperturbed value $\Trace[\hat{W}\hat{\rho}^0]$ for the average work $\work$, then the QNDM quasi-probability density function of the work exhibits at least a negative region. Specifically, for a work distribution, these negative regions are associated with non-classical energy exchanges. For example, consider a two-level system with bare Hamiltonian $\hat{H}=(\hbar\omega/2)\hat{\sigma}_z$, where $\hat{\sigma}_z$ is the Pauli matrix along the $z$-axis. The system is driven by a cyclic transformation. This means that the energy spectrum is the same at the beginning and at the end of the system time evolution, i.e., the energies $\epsilon_j(\Time)$ and $\epsilon_j(0)$ ($j=1,2$) take the same values $\pm \hbar\omega/2$. The TPM scheme --- providing the classical contribution to $\Prob_{\rm QNDM}(W)$ --- allows only for energy exchanges equal to $0$ and $\pm\hbar\omega$. On the contrary, due to the presence of quantum coherence in the initial density operators $\hat{\rho}^0$, the work distribution develops quantum contributions, and the support of $\Prob_{\rm QNDM}(W)$ comprises the work realizations $\pm \hbar\omega/2$, $0$, and $\pm \hbar\omega$. The `fractional' energy exchanges $\pm \hbar\omega/2$ are not allowed by any classical and statistical interpretation, and cannot be recovered by the TPM scheme.

We point out that the experimental protocol consists of measuring experimentally the quasi-characteristic function $\G$ such that $\Prob_{\rm QNDM}(W)$ is reconstructed from the analysis of the experimental data. In addition, we observe that the first and second moments of the QNDM work quasi-probability density function have a clear physical interpretation, extending the validity of these concepts from classical thermodynamics. However, the structure (in particular, the time-ordering) of the higher statistical moments may not be physically motivated. The latter, indeed, are ultimately determined and cannot be separated by the measurement procedure.

\subsection{Non-demolition measurements for open quantum system}
\label{sec:open_system}

In this section, we briefly discuss the application of the QNDM protocol to open quantum systems~\cite{petruccione2002theory}. We will then make use of these concepts in Sec.~\ref{sec:quantum_to_classical}, where they are contextualized for the tracking of the quantum-to-classical transition. In order to keep the presentation affordable, here we just argue the main ideas and advantages of applying QNDM to monitor the dissipated heat in coherently-driven and open dynamics due to system-environment interactions.

The dissipated heat is a process function, and thus depends on the full time evolution of the system. For example, a quantum system starting and ending in the ground state does not change its overall internal energy. However, due to the interaction with the external environment, the system can produce a possibly arbitrary amount of heat during the open dynamics, through a sequence of excitations and relaxations. The only way to keep track of the dissipated heat is by constantly monitoring the dynamics of the system.

When it comes to the dissipated heat at the quantum level, a standard approach~\cite{campisi2011colloquium} is to perform sequential measurements of the {\it environmental degrees of freedom}, in order to evaluate the energy dissipated into the environment, which is identified as heat. However, this approach may be both conceptually and practically unfeasible.\\ 
First, in many applications, a nanoscale system couples with an environment that has to be considered macroscopic, namely possessing a virtually infinite number of degrees of freedom and an essentially infinite heat capacitance. This entails a common approximation (Born's approximation)~\cite{petruccione2002theory, Blum_book} in open quantum systems, i.e., to assume that the state of the environment does not change because of the interaction with the nanoscale system. As a result, from a practical point of view, one would need to detect any absorbed/emitted energy quantum from the environment, or at least to statistically sample a large portion of the environmental degrees of freedom.\\
Second, an environment with a small number of degrees of freedom changes its state because of the absorption or emission of an energy quantum, and it behaves quantum mechanically. These energy exchanges would lead to a non-equilibrium environment (e.g., an environment without a definite temperature), so it must be treated as an additional evolving quantum system. In addition, a measurement on it induces a wave-function collapse, destroying (partially or totally) its properties.

These conditions are so restrictive that only a few proposals have been able to meet all the requirements to fulfil them. For example, in Ref.~\cite{pekola2013calorimetric,Gasparinetti2015,Viisanen_2015,Saira2016,Vesterinen2017,Ronzani2018}, the use of an auxiliary system is proposed to collect the dissipated energy, under the assumption that the auxiliary system has a quantized spectrum and at the same time enough degrees of freedom to act as an intermediate, effective environment. However, despite these rare situations, it seems questionable whether the proposal to measure the environmental energy can always be exploited efficiently.

The QNDM approach can overcome these drawbacks for monitoring the heat, because it is designed to extract information from the {\it system's degrees of freedom}, without inducing any perturbation on the dynamics. The important requirement to use the QNDM approach for monitoring the heat over time is that the time-scales of the external drive (performing work on the system) and the environment are separated~\cite{solinas2015fulldistribution}.
Under this assumption, we can exploit~\cite{solinas2013work} the result stating that time-changes of the system's Hamiltonian are associated with work, while time-changes in the density operator of the system are associated with dissipated heat~\cite{Alicki1979}. Hence, performing fast system-detector couplings, the monitoring of the system's internal energy changes leads to reconstructing energy dissipation in terms of heat. Then, over a longer time scale, the effects due to the external drive in terms of work can also be included, with the constraint of respecting the first law of thermodynamics~\cite{solinas2015fulldistribution}.

\section{Certification of quantum and classical behaviours}
\label{sec:macrorealism_and_LGI}

We are going to show that the negativity of the QNDM quasi-probability density function is a necessary and sufficient condition for the identification of quantum superpositions in the system's state at the measurement times. Accordingly, given an initial system's state and a measurement observable, determining a negative QNDM quasi-probability certifies whether the system under scrutiny exhibits quantum behaviours.

Furthermore, we will show that if the QNDM quasi-probability density function turns out to be positive, then the system has classical properties as intended in Ref.~\cite{GherardiniTutorial};
that is, the system's states at the measurement times $t_0, t_1, t_2$ are necessarily diagonal with respect to the basis that spectrally decomposes the measurement observable. The latter condition entails that the QNDM quasi-probabilities can be provided by the TPM scheme (sequential projective measurements, operationally), meaning that it is identically equal to a classical probability distribution.

In order to effectively illustrate these findings, we discuss the prototypical case of three non-demolition measurements that allows us to compare the certification of quantum behaviours obtained from the QNDM and the standard LGIs.

\subsection{Three-measurement scheme}
\label{sec:3measurement}

The simplest and most used version of the Leggett-Garg inequalities (LGIs)~\cite{Leggett1985} is based on a three measurement scheme~\cite{Emary_2014,Halliwell2016}.
To discuss the relations between the QNDM protocol and LGIs, we extend the protocol discussed in Sec.~\ref{sec:QNDM_2measurement} to the three-measurement scheme depicted in Fig.~\ref{fig:superpositions}. For technical details, the reader can refer to Ref.~\cite{solinas2024}.

Let us thus consider a system that evolves under a unitary transformation in the time interval $0 \leq t \leq \mathcal{T}$, and the system observable $\hat{A}$ is measured sequentially at times $t_0\leq t_1 \leq t_2=\mathcal{T}$. As above, the measurements of $\hat{A}$ are performed by coupling the system with a detector that interacts through the operator $\hat{u}$. The system's time evolution in between the measurements is governed by the unitary operators $\MU_k \equiv \MU( t_k, t_{k-1} )\otimes \idop$ with $k=1,2$. Hence, overall, the time evolution operator $\MU_{tot}$ for the system-detector within the interval $[t_0,t_2]$ read as
\begin{equation}
\MU_{tot} = \hat{u} \MU_2 \hat{u} \MU_1 \hat{u} = e^{ i \frac{\lambda}{2} \hat{A} \otimes \hat{p} } \MU_2 \, e^{ i \frac{\lambda}{2} \hat{A} \otimes \hat{p} } \MU_1 \, e^{ i \frac{\lambda}{2} \hat{A} \otimes \hat{p} }.
\label{eq:U_tot_3meas}
\end{equation}
Assuming same initial states, notations and definitions as in Sec.~\ref{sec:QNDM_2measurement}, the QNDM quasi-characteristic function is \cite{solinas2024}
\begin{equation}\label{eq:quasi_char_QNDM}
    \G = \sum_{i, l, j, m, k} e^{i \lambda \left (a_k + \frac{a_j+a_m+a_i+a_l}{2} \right)}   U_{2,kj} U_{1,ji} \, \rho^0_{il} U^{*}_{1,lm} U^{*}_{2,mk}\,,
\end{equation}
where 
\begin{equation}\label{eq:stochastic_Delta}
    \Delta_{k, j, m, i, l} := a_k + \frac{ a_j+a_m+a_i+a_l }{2}
\end{equation}
denotes the spectrum of measurement outcomes at times $t_0,t_1,t_2$.

By direct calculation of the inverse Fourier transform of the QNDM quasi-characteristic function (\ref{eq:quasi_char_QNDM}), we obtain the quasi-probability density function $\Prob_{\rm QNDM}(\Delta)$ that is formally equal to
\begin{equation}\label{eq:ND_distribution_2}
    \Prob_{\rm QNDM}(\Delta) = \sum_{i, l, j, m, k} P_{\rm QNDM}(k, j, m, i, l) \, \delta\left[ \Delta - \Delta_{k, j, m, i, l} \right].
\end{equation} 
In Eq.~\eqref{eq:ND_distribution_2},
\begin{equation}
P_{\rm QNDM}(k, j, m, i, l) = U_{2,kj} U_{1,ji} \, \rho^0_{il} U^{*}_{1,lm} U^{*}_{2,mk}={\rm Tr}\left[ \hat{\Pi}_k \MU_2 \hat{\Pi}_j \MU_1 \hat{\Pi}_i \, \hat{\rho}^0 \hat{\Pi}_l \MU^\dagger_1 \hat{\Pi}_m \MU^\dagger_2 \right]
\end{equation}
are the interference contributions for the system to arrive in the state $\ket{k}$ following the superposition of paths $i \rightarrow j \rightarrow k$ and $l \rightarrow m \rightarrow k$ (see Fig.~\ref{fig:superpositions}).

Again, we can separate the classical $\Prob_{cl}(\Delta)$ from the quantum $\Prob_{q}(\Delta)$ contributions of $\Prob_{\rm QNDM}(\Delta)$. The first ones are obtained when $m=j$ and $l=i$, i.e.,
\begin{equation}\label{eq:Pcl_Delta}
    \Prob_{cl}(\Delta) = \sum_{i, j, k} P_{\rm QNDM}(k, j, j, i, i) \, \delta\left[ \Delta - \Delta_{k, j, j, i, i}   \right]
\end{equation} 
that comprise positive probabilities. This is because $P_{\rm QNDM}(k, j, j, i, i) = P^{(0)}_{i} P^{(1)}_{i \rightarrow j} P^{(2)}_{j \rightarrow k}$, where $P^{(0)}_i = \rho^0_{ii}$, and $P^{(1)}_{i \rightarrow j} = |U_{1,ji}|^2, P^{(2)}_{j \rightarrow k} = |U_{2,kj}|^2$ are the probabilities the system is initially in the state $\ket{i}$ and then makes a transition from state $\ket{i}$ to $\ket{j}$ and from $\ket{j}$ to $\ket{k}$, respectively.

The quantum contributions $\Prob_{q}(\Delta)$ of $\Prob_{\rm QNDM}(\Delta)$, instead, are present when $m\neq j$ or $l\neq i$. Thus, they are associated with the coherent superposition of paths or trajectories in the space identified by the sequence of measurement outcomes~\cite{solinas2013work, solinas2024}. They formally read as 
\begin{equation}
     \Prob_{q}(\Delta) = \sideset{}{'}\sum_{i, l, j, m, k} P_{\rm QNDM}(k, j, m, i, l) \, \delta \left[ \Delta - \Delta_{k, j, m, i, l}  \right], 
     \label{eq:P_q}
\end{equation}
where the sum $\sideset{}{'}\sum$ includes contributions with ($l = i$, $m \neq j$), ($m = j$, $l \neq i$) and ($m \neq j$, $l \neq i$). These are the conditions under which the system is in a superposition of the $\hat{A}$'s eigenstates, in correspondence with at least one measurement in the sequence.

\subsubsection{QNDM distribution implies unperturbed marginals}
\label{sec:NIM}

The quasi-probability density function in \eqref{eq:ND_distribution_2} has a significant mathematical property: the marginals of $\Prob_{\rm QNDM}(\Delta)$, over all the measurement outcomes but the ones at a single time, return the `unperturbed' probability to measure the observable $\hat{A}$ given by the Born's rule, without any intermediate measurement of the sequence affects the result.

To see this in the three-measurement scheme, the probability $P(a_k)$ of measuring the outcome $a_k$ at time $t_2$ is computed by summing the QNDM quasi-probability $P_{\rm QNDM}(k, j, m, i, l)$ over all the intermediate outcomes $a_i$, $a_l$, $a_j$, $a_m$. Using the properties of the projectors, i.e., $\hat{\Pi}_k^2 = \hat{\Pi}_k$, $\sum_k \hat{\Pi}_k = \idop$ and $\hat{\Pi}_k \hat{\Pi}_j = \delta_{kj}\hat{\Pi}_k$, we find:
\begin{equation}
    P(a_k) = \sum_{i, l, j, m} P_{ND}(k, j, m, i, l) =  \sum_{i, l, j, m} \Trace \left[ \hat{\Pi}_k \MU_2 \hat{\Pi}_j \MU_1 \hat{\Pi}_i \, \hat{\rho}^0 \hat{\Pi}_l \MU^\dagger_1 \hat{\Pi}_m \MU^\dagger_2 \right] = \Trace \left[ \hat{\Pi}_k \MU \hat{\rho}^0 \MU^\dagger \right]  = \Trace \left[ \hat{\Pi}_k \, \hat{\rho}(t_2) \right],  
\end{equation}
where $\hat{U} = \MU_2 \MU_1$ and $\hat{\rho}(t_2) = \MU \hat{\rho}^0 \MU^\dagger$. Therefore, $P(a_k)$ is the probability to record the measurement outcome $a_k$ at time $t_2$ as given by Born's rule~\cite{Halliwell2016}. We also compute the single-time probabilities $P(a_i)$, $P(a_l)$ for the initial measurement at time $t_0=0$:
\begin{eqnarray}
    P(a_i) &=& \frac{1}{2}\sum_{l, j, m, k} \Big\{ P_{ND}(k, j, m, i, l) + P_{ND}(k, j, m, l, i) \Big\} =  \nonumber \\
    &=& \frac{1}{2} \sum_{l, j, m, k} \Big\{  \Trace \left[ \hat{\Pi}_k \MU_2 \hat{\Pi}_j \MU_1 \hat{\Pi}_i \, \hat{\rho}^0 \hat{\Pi}_l \MU^\dagger_1 \hat{\Pi}_m \MU^\dagger_2 \right]  + \Trace \left[ \hat{\Pi}_k \MU_2 \hat{\Pi}_j \MU_1 \hat{\Pi}_l \, \hat{\rho}^0 \hat{\Pi}_i \MU^\dagger_1 \hat{\Pi}_m \MU^\dagger_2 \right] \Big\} = \Trace \left[ \hat{\Pi}_i \, \hat{\rho}^0 \right] 
\end{eqnarray}
and
\begin{equation}
P(a_l) = \frac{1}{2}\sum_{i, j, m, k} \Big\{ P_{ND}(k, j, m, i, l) + P_{ND}(k, j, m, l, i) \Big\} = \Trace \left[ \hat{\Pi}_l \, \hat{\rho}^0 \right];
\end{equation}
both these probabilities satisfy Born's rule, as well as $P(a_k)$. Moreover, for the measurement at time $t_1$, there are two possible realizations over the indices $j$ and $m$, so that 
\begin{equation}
    P(a_j) = \frac{1}{2}\sum_{i, l, m, k} \Big\{  P_{ND}(k, j, m, i, l) + P_{ND}(k, m, j, i, l) \Big\} = \Trace \left[ \hat{\Pi}_j \MU_1 \hat{\rho}^0  \MU^\dagger_1 \right] = \Trace \left[ \hat{\Pi}_j \, \hat{\rho}(t_1) \right],
\end{equation}
which is the probability of recording the outcome $a_j$ after the system has evolved to the state $\hat{\rho}(t_1) = \MU_1 \, \hat{\rho}^0  \MU^\dagger_1$, without the initial measurement at time $t=0$ has affected the system's state.

Finally, summing the QNDM quasi-probabilities $P_{\rm QNDM}(k, j, m, i, l)$ over one of the possible three-index sequences returns the Kirkwood-Dirac quasi-probabilities at two-times~\cite{LostaglioKirkwood2022,GherardiniTutorial,ArvidssonShukur2024review}, apart from the 3-tuples $(j,m,k)$ and $(i,l,k)$ that give rise to a single-time probability. By definition, also Kirkwood-Dirac quasi-probabilities return marginal distributions that are not perturbed by any intermediate measurement performed previously.

\subsection{Theoretical basis for quantumness and classicality certification}
\label{sec:necessary_and_sufficient}

With a few exceptions of negligible statistical significance (corresponding to isolated points in a large parameter space) that can be cured as discussed in the next section \ref{sec:quantumness_certification}, we can prove that $\Prob_{q}(\Delta) \neq 0$ always identifies whether the system's state is in a superposition of the $\hat{A}$'s eigenstates.

The {\it quantumness certification} stems from the following statement: the presence of negative regions in the QNDM quasi-probability density function $\Prob_{\rm QNDM}(\Delta)$ is a necessary and sufficient condition for the violation of the MRps~\cite{solinas2024}. In other words, {\it a system is in a coherent superposition of the measured observable's eigenstates if and only if $\Prob_{\rm QNDM}(\Delta)$ exhibits negativity, i.e., it has negative contributions}. The fact that the measured system is in a superposition state violates the MRps and, consequently, the MR.

This result is based on the following three main properties of the QNDM protocol, derived in Ref.~\cite{solinas2024}. First, the QNDM quasi-probability density function $\Prob_{\rm QNDM}(\Delta)$ comprises only real numbers. Second, $\Prob_{\rm QNDM}(\Delta)$ is normalized to $1$, and this normalization comes uniquely from the classical contribution $\Prob_{cl}(\Delta)$. The first and second properties imply that the terms entering the quantum contribution $\Prob_{q}(\Delta)$ of $\Prob_{\rm QNDM}(\Delta)$ cancel out each other, such that some of them must be negative.

The negativity of $\Prob_{\rm QNDM}(\Delta)$ appears whenever the system is in a superposition of the measurement observable $\hat{A}$'s eigenstates, in at least one measurement time among $t_0, t_1, t_2$. This is because the negativity of $\Prob_{\rm QNDM}(\Delta)$ univocally means that $\Prob_{q}(\Delta) \neq 0$, which occurs {\it if and only if} the system's state is in a superposition of the $\hat{A}$'s eigenstates. Hence, since the integral of $\Prob_{q}(\Delta)$ over all values of $\Delta$ is equal to zero ($\int d\Delta \Prob_{q}(\Delta) = 0$), $\Prob_{\rm QNDM}(\Delta)$ must contain negative real terms whenever $\Prob_{q}(\Delta)\neq 0$.

The positivity of the QNDM quasi-probability density function gives us complementary information about the system's behaviours, and can thus be used as a witness for {\it classicality certification}. In fact, since the positivity of the QNDM distribution $\Prob_{\rm QNDM}(\Delta)$ implies the absence of quantum superposition, it entails that the system's states at the measurement times $t_0, t_1, t_2$ are diagonal along the basis that spectrally decomposes the observable $\hat{A}$ at such times~\cite{GherardiniTutorial}. Hence, $\Prob_{q}(\Delta) = 0$ and $\Prob_{\rm QNDM}(\Delta)=\Prob_{\rm cl}(\Delta)$. As discussed in the previous section, $\Prob_{\rm cl}(\Delta)$ is obtained from implementing the TPM scheme~\cite{GherardiniTutorial}, which is realized by a sequence of projective measurements of $\hat{A}$ at times $t_0, t_1, t_2$. In this sense, $\Prob_{\rm cl}(\Delta)$ is denoted as being classical. In the conditions where the system's states are diagonal from the positivity of the QNDM distribution, the TPM protocol does not entail any perturbation of the outcomes' statistics from measuring $\hat{A}$. This means that the TPM distribution suffices to describe the statistics of measurement outcomes at times $t_0, t_1, t_2$, as it is able to return the unperturbed value of the corresponding statistical moments~\cite{GherardiniTutorial}. The unperturbed moments are obtained because the system's states are diagonal along the observable's basis at the measurement times. Therefore, no quantum coherence and/or correlation is erased on such a basis, meaning that a classical path must be followed (red dashed curve in Fig.~\ref{fig:superpositions}).

Alternatively, the concepts about classicality certification discussed in the latter paragraph can be figured out by noting that each term $P_{\rm QNDM}(k, j, j, i, i)$ comprising the positive distribution $\Prob_{\rm cl}(\Delta) = \sum_{i, j, k} P_{\rm QNDM}(k, j, j, i, i) \, \delta[ \Delta - \Delta_{k, j, j, i, i}]$ in \eqref{eq:Pcl_Delta} is associated with the averages and correlators $\average{\hat{A}_i} = \sum_{a_i} a_i\,P_i^{(0)}(a_i)$, $\average{\hat{A}_j}= \sum_{a_i,a_j} a_j\,P_i^{(0)}(a_i) \, P_{i\to j}^{(1)}(a_j | a_i)$, $\average{\hat{A}_i \hat{A}_j}  
= \sum_{a_i,a_j} a_i a_j \,
P_i^{(0)}(a_i) \, P_{i\to j}^{(1)}(a_j | a_i)$, 
$\average{\hat{A}_i \hat{A}_j \hat{A}_k}
= \sum_{a_i,a_j,a_k} a_i a_j a_k\,
P_i^{(0)}(a_i) \, P_{i\to j}^{(1)}(a_j | a_i) \, P_{j\to k}^{(2)}(a_k | a_j)$ and so on, built over the classical path $i \rightarrow j \rightarrow k$. In particular, 
\begin{equation}
P_{\rm QNDM}(k, j, j, i, i) = \frac{1}{8}\Big[1 + a_i \average{\hat{A}_i}  + a_j \average{\hat{A}_j} + a_k \average{\hat{A}_k} +  a_i a_j \average{\hat{A}_i \hat{A}_j} +  a_i a_k \average{\hat{A}_i \hat{A}_k} +
+ a_j a_k \average{\hat{A}_j \hat{A}_k} + a_i a_j a_k \average{\hat{A}_i \hat{A}_j \hat{A}_k} \Big]
\end{equation}
that is in the same form as a generic three-time probabilities as discussed in Ref.~\cite{HalliwellConference2019}. Therefore, we can conclude that the positivity of $P_{\rm QNDM}(k, j, j, i, i)$ can be used to identify the classical behaviour of the quantum system.

\subsection{Operational procedure for quantumness certification}
\label{sec:quantumness_certification}

At this point, we can show how the QNDM quasi-probability density function can be extracted with a feasible and operational procedure that always guarantees the certification of quantum features in a system's state. To do this, we explicitly take into account the few exceptions whereby $\Prob_{q}(\Delta)$ is not able to identify superpositions of the $\hat{A}$'s eigenstates in the system's state.

First, by construction, $\Prob_{q}(\Delta)$ comprises real numbers that are the real parts of $P_{\rm QNDM}(k, j, m, i, l)$, complex numbers. This means that the sum of $P_{\rm QNDM}(k, j, m, i, l)$ could vanish even if the system is a coherent superposition of $\hat{A}$'s eigenstates, when the quasi-probabilities $P_{\rm QNDM}(k, j, m, i, l)$ are purely imaginary. To avoid this, it is sufficient to run the measurement protocol two more times, by performing the additional unitary transformation $\hat{U}_{\chi}:= \exp\{ i \chi \hat{A} \}$ on the system, first at time $t_0$ and then at time $t_1$ in the second run. The transformation $\hat{U}_{\chi}$ just adds a phase in the system's state and does not induce transitions between the eigenstates of the observable $\hat{A}$. Therefore, it does not change the fact that the system is in a superposition state. A proper choice of $\chi$ (even sampled by a probability distribution could be a good option) ensures that every pure imaginary contribution encoded in $P_{\rm QNDM}(k, j, m, i, l)$ acquires a real part, which results in $\Prob_{q}(\Delta) \neq 0$.

In addition, the whole procedure for quantumness certification becomes more {\it robust} if the second and third measurements at times $t_1$ and $t_2$ are made over the measurement observable with slightly shifted coupling parameters: $\lambda \rightarrow \lambda' = \lambda + \delta$ and $\lambda \rightarrow \lambda'' = \lambda + \delta'$. These measurements are realized by applying the unitary transformation $\hat{u}_{\delta} := e^{ i \frac{\lambda'}{2} \hat{A} \otimes \hat{p} }$ and $\hat{u}_{\delta'} := e^{ i \frac{\lambda''}{2} \hat{A} \otimes \hat{p} }$ to the system-detector. 
The advantages of using $\hat{u}_{\delta}, \hat{u}_{\delta'}$ are to split the support of $\Prob_{\rm QNDM}(\Delta)$ in $\Delta$, and to remove the degeneracy among the different paths associated with the same value of $\Delta_{k, j, m, i, l}$. For example, for $i=j$, $l=m$, and $i \neq l$, the system is in a coherent superposition of states and $\Delta=a_k$. However, the same value of $\Delta$ is also associated with the classical path with $i=l$ and $j=m$. This fact could happen for binary observables when ($a_i=a_j=1,\,a_m=a_l=-1$) or ($a_i=a_j=-1,\,a_m=a_l=1$). But if we split the support of $\Prob_{\rm QNDM}(\Delta)$, we can guarantee that both classical and quantum contributions are present in $\Delta$.

Overall, the procedure for certifying a violation of MRps (and thus of MR) is given by the following three runs of experiments:
\begin{eqnarray*}
    && (1) \quad \MU_{tot} = e^{ i \frac{\lambda^{\prime \prime}}{2} \hat{A} \otimes \hat{p} } \MU_2 \, e^{ i \frac{\lambda^\prime}{2} \hat{A} \otimes \hat{p} } \MU_1 \, e^{ i \frac{\lambda}{2} \hat{A} \otimes \hat{p} } \\
    && (2) \quad \MU_{tot} = e^{ i \frac{\lambda^{\prime \prime}}{2} \hat{A} \otimes \hat{p} } \MU_2 \, e^{ i \frac{\lambda^\prime}{2} \hat{A} \otimes \hat{p} } \MU_1 \, e^{ i \frac{\lambda}{2} \hat{A} \otimes \hat{p} } e^{ i \chi \hat{A} \otimes \idop }\\
    && (3) \quad \MU_{tot} = e^{ i \frac{\lambda^{\prime \prime}}{2} \hat{A} \otimes \hat{p} } \MU_2 \, e^{ i \frac{\lambda^\prime}{2} \hat{A} \otimes \hat{p} } e^{ i \chi \hat{A} \otimes \idop } \MU_1 \, e^{ i \frac{\lambda}{2} \hat{A} \otimes \hat{p} } .
\end{eqnarray*}
If none of the resulting QNDM quasi-probability density functions $\Prob_{\rm QNDM}(\Delta)$ show negative regions, then MRps is fulfilled.

\subsection{Connection and comparison with Leggett-Garg inequalities}
\label{sec:QNDM_vs_LGIs}

In this section, we compare the QNDM protocol with the usual tool employed so far to identify the violation of MRps, i.e., the Leggett-Garg inequalities. The LGIs are usually formulated for binary observables~\cite{Leggett1985,Fritz_2010,Emary_2014,Halliwell2016}, namely for observables that can have two outcomes. Hence, we now assume that the eigenvalues of the observable $\hat{A}$ are $a(t_n) = \pm 1$ for any time $t_n$. LGIs are built using quantum correlators at two times, say $t_n$ and $t_m$, which are defined as~\cite{Fritz_2010,Emary_2014,Halliwell2016} 
\begin{equation}
C_{nm} = \frac{1}{2}{\rm Tr}\Big[ \big(\hat{A}^{H}(t_n) \hat{A}^{H}(t_m) + \hat{A}^{H}(t_m) \hat{A}^{H}(t_n) \big) \hat{\rho}^0\Big]
\end{equation}
with $\hat{A}^{H}(t) = \MU^\dagger(t, 0)\hat{A}\MU(t, 0)$ represented in Heisenberg picture. The expressions of $C_{nm}$ with $n,m=0,1,2$ and $n<m$ are respectively equal to
\begin{eqnarray}
    C_{01} &=& \sum_{i, j, k} a_i a_j P(a_k,a_j,a_i)\\
    C_{12} &=& \sum_{i, j, k} a_j a_k P(a_k,a_j,a_i)\\
    C_{02} &=& \sum_{i, j, k} a_i a_k P(a_k,a_j,a_i),
\end{eqnarray}
where $P(a_k, a_j, a_i) = P^{(0)}_{i} P^{(1)}_{i \rightarrow j} P^{(2)}_{j \rightarrow k} = P_{\rm QNDM}(k, j, j, i, i)$ is the joint probability to record the measurement outcomes $a_i, a_j, a_k$ at times $t_0,t_1,t_2$ using a sequential scheme of projective measurements. Following Ref.~\cite{Emary_2014}, we introduce the Leggett-Garg (LG) parameter 
\begin{equation}\label{eq:def_LG_param}
K := C_{01} + C_{12} - C_{02}
\end{equation}
that reads as $K = 1 - 4\left[ P(1,-1,1) + P(-1, 1,-1) \right]$ and obeys the LGI $-3 \leq K \leq 1$ if the system under consideration is classical and fulfils all the assumptions of MR. This means that, if the $-3 \leq K \leq 1$ is not satisfied, then at least one condition of MR is violated. However, the opposite is not guaranteed because there are cases where the MR is violated, but the value of $K$ is still included in the range $-3 \leq K \leq 1$.

Starting from the QNDM quasi-probability density function \eqref{eq:ND_distribution_2}, for binary observables the correlators $C_{nm}$ can be equivalently expressed as~\cite{Halliwell2016,solinas2024} 
\begin{equation}\label{eq:C_ij}
C_{nm} = \sum_{u,v} a_{u_n} a_{v_m} P_{\rm QNDM}(v_m, u_n),
\end{equation}
where $a_{u_n} := a_u(t_n)$, $a_{v_m} := a_v(t_m)$, and $P_{\rm QNDM}(v_m, u_n)$ are Kirkwood-Dirac quasi-probabilities~\cite{LostaglioKirkwood2022,GherardiniTutorial,ArvidssonShukur2024review}, as commented in Sec.~\ref{sec:NIM}. For example, $C_{01} = \sum_{i,j}a_i a_j P_{\rm QNDM}(j, i)$ with
\begin{equation}
    P_{\rm QNDM}(j, i) = \sum_{l, m, k}\left[ P_{\rm QNDM}(k, j, m, i, l) +  P_{\rm QNDM}(k, m, j, i, l) + P_{\rm QNDM}(k, j, m, l, i) +  P_{\rm QNDM} (k, m, j, l, i) \right].
\end{equation}
From Eq.~\eqref{eq:C_ij}, we thus evince that the LG parameter $K$ can be written as a function of the QNDM quasi-probabilities at three times, whose distribution implicitly contains the information of the two-times correlators $C_{nm}$ and of the three-time correlator $C_{ijk}=\sum_{i,j,k}a_i a_j a_k P(a_k, a_j, a_i)$. Accordingly, using the QNDM quasi-probability density function, one can directly single out the value of $K$ and understand whether the LGI is violated.

We now report and comment on the expression of the LG parameter $K$ as a function of the QNDM quasi-probabilities at three times; the derivation of this result is in Ref.~\cite{solinas2024}. Since we are working with binary variables, each of them can take two values: $a_n$ and $-a_n$. For simplicity, we refer to them with the indices $n$ and $\bar{n}$; for example, $P_{\rm QNDM}(\bar{k}, j, m, i, \bar{i})$ stands for $P_{\rm QNDM}(-a_k, a_j, a_m, a_i, -a_i)$. Using Eq.~\eqref{eq:C_ij}, the LG parameter $K$ of Eq.~\eqref{eq:def_LG_param} reads as
\begin{equation}\label{eq:K_ND}
    K =  \frac{1}{4} \sum_{i, l, j, m, k} f(k, j, m, i, l) P_{\rm QNDM}(k, j, m, i, l)
\end{equation}
with $f(k, j, m, i, l) = (a_i +a_l)(a_j +a_m) + a_k (a_j +a_m-a_i -a_l)$. It is convenient to separate $K$ in its classical and quantum contributions: $K = \KND_{cl} + \KND_{q,1} + \KND_{q,2}$. The classical contribution $\KND_{cl}$ is obtained when $(m=j, l=i)$, while the quantum ones $\KND_{q,1}$ and $\KND_{q,2}$ correspond to $(m=j, l=\bar{i} \neq i)$ and $(m=\bar{j} \neq j, l= i)$, respectively. It can be shown~\cite{solinas2024} that $\KND_{cl}$ satisfies the LGI, i.e., $-3 \leq \KND_{cl} \leq 1$, and that $\KND_{q,1}=0$.
Therefore, when present, the violation of the LGI is due to the second quantum contribution $\KND_{q,2}$ of $K$, which is equal to
\begin{equation}\label{eq:final_K_q2}
    \KND_{q,2} =-4 \sum_{k} \re\left[ P_{\rm QNDM}(k, j, \bar{j}, k, k) \right].
\end{equation}
The contributions with $m = \bar{j} \neq j$ and $l=i$ entering $\KND_{q,2}$ are entailed by some superpositions between eigenstates of $\hat{A}$, thus violating MRps. This implies, in turn, that whenever the MRps is violated but the LGI is satisfied, we are dealing with a combination of initial state, system dynamics, and measurement observable such that $\KND_{q,2}$ vanishes. In such a case, the total quantum contribution $\KND_{q}$ vanishes as well, $\KND_{q} = 0$, with the result that $\KND = \KND_{cl}$, confirming that the violation of the LGI is only a sufficient condition for identifying violations of MRps. Indeed, there could be trajectories in which $l \neq i$ (thus with a superposition of $\hat{A}$'s eigenstates) that are not identified by LGI since $\KND_{q,1}=0$. On the contrary, if we take into account $\Prob_{\rm QNDM}(\Delta)$, the terms $P_{\rm QNDM}(k, j, j, i, \bar{i})$ in \eqref{eq:ND_distribution_2} always contribute to negative regions of the distribution, whose presence is a sufficient and necessary condition for the violation of MRps.

Overall, the QNDM approach does not have to face with constraints on the choice of parameters, initial state, and measurement times, as it occurs for standard methods in the literature~\cite{Clemente2015,ClementePRL2016,Halliwell2017,HalliwellConference2019} that address the problem of detecting the presence of quantum superposition states, in the initial system's state and/or generated dynamically. From an operational standpoint, QNDM can thus represent an efficient and robust protocol for certifying quantumness, albeit at the cost of using an auxiliary system that acts as a detector.

\begin{figure}
   \centering
   \includegraphics[scale=.6]{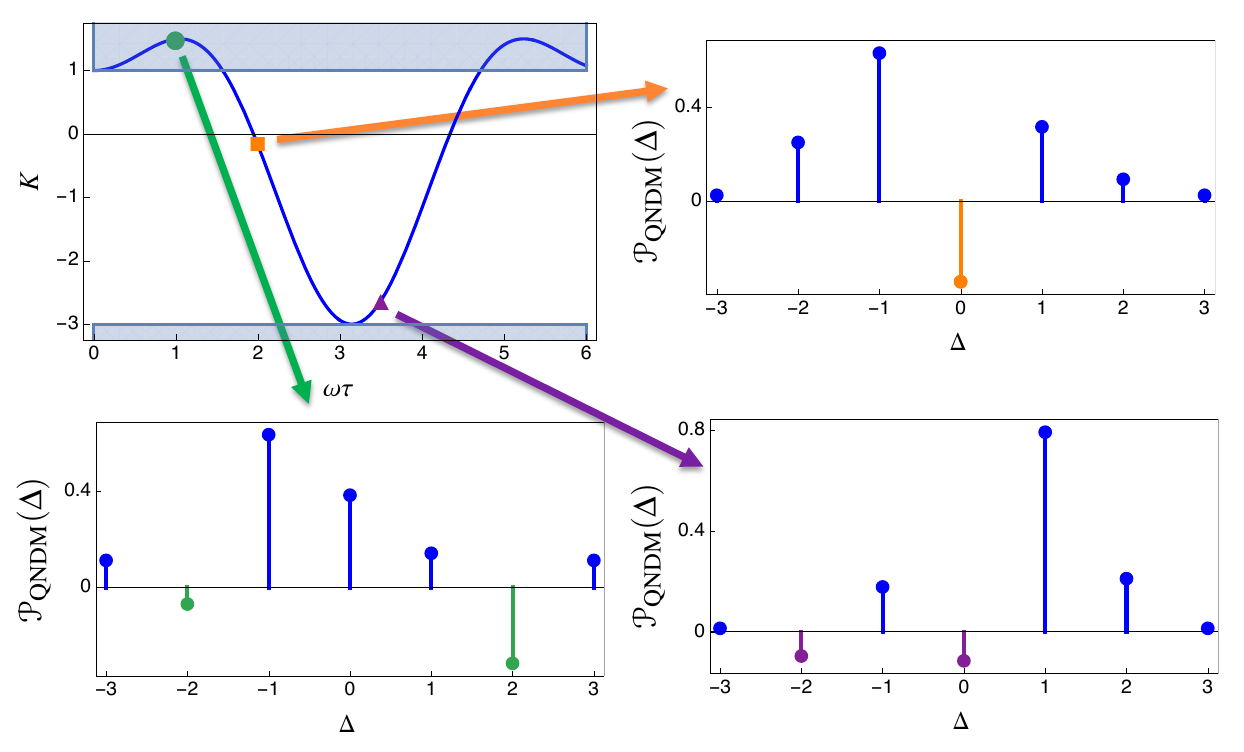}
   \caption{
   Comparison between the prediction of the LGI and QNDM approach. In the top-left panel, we plot the LG parameter $K$ as a function of $\omega\tau$. The shaded regions are the ones for which the LGI is not satisfied. The other panels show the quasi-probability distribution $\Prob_{\rm QNDM}(\Delta)$ evaluated at three different values of $\omega\tau$. The distribution $\Prob_{\rm QNDM}(\Delta)$ always shows negative regions, while only for the green circle point the LGI is violated.
   } 
   \label{fig:LG_VS_PD}
\end{figure} 
\subsubsection{Example}
\label{example_LGI}

Let us consider a two-level quantum system that is initialized in the state $\ket{\psi_0} = (\ket{0} + i \ket{1})/\sqrt{2}$ where $\ket{0}$ and $\ket{1}$ are the eigenstates of the Pauli operator $\hat{\sigma}_z$. The dynamics is generated by the Hamiltonian $\hat{H} = \omega\hat{\sigma}_x/2$, and we measure the operator $\hat{A}=\hat{\sigma}_z$ at times $t_0=0$, $t_1=\tau$ and $t_2=2\tau$~\cite{Halliwell2016, solinas2024}.

The two-times correlators $C_{nm}$ and the corresponding LG parameter $K$ can be calculated analytically, such that $K = 2 \cos(\omega\tau) - \cos(2 \omega \tau)$. As discussed above, any value of $K$ outside the range $[-3,1]$ implies the violation of at least one MR condition; instead, it is not guaranteed that the MR holds if $-3 \leq K \leq 1$.

The numerical results for the example are shown in Fig.~\ref{fig:LG_VS_PD}. 
The top-left panel shows the value of $K$ as a function of $\omega \tau$ with the shaded regions representing the situations in which the LGI is not satisfied. The colored shapes (dot, square, triangle) give the value of $K$ for three specific values of $\omega\tau$. The other panels of Fig.~\ref{fig:LG_VS_PD} depict the QNDM quasi-probability distribution associated with each colored shape. In correspondence with the green circle, the LGI is violated, and the corresponding QNDM distribution $\Prob_{\rm QNDM}(\Delta)$ has negative regions. However, for the other two colored shapes (the orange square and the purple circle), a more interesting fact occurs: while the LGI is satisfied, $\Prob_{\rm QNDM}(\Delta)$ presents negative regions, and it allows us to identify quantum features of the system even when the LGI fails. From the analytical expression of $K$, the violation of the LGI occurs for $0 \leq \omega \tau \leq \pi/2$ and $3 \pi/2 \leq \omega \tau \leq 2 \pi$. Hence, in this example, the LGI correctly identifies the violation of MRps in only half of the cases, while the QNDM protocol always succeeds.

\section{Resources for realizing QNDM, and robustness against environmental noise}
\label{sec:QNDM_robustness}

As discussed in Secs.~\ref{sec:necessary_and_sufficient} and \ref{sec:QNDM_vs_LGIs}, the QNDM approach allows for identifying the presence of quantum features in a system's state even when the LGI fails. It is thus natural to ask how many resources one needs to realize a sequence of non-demolition measurements, and compare them to the resources required to reconstruct the LGI. Here, we refer to the protocols discussed in Sec.~\ref{sec:QNDM_vs_LGIs}, including the example in \ref{example_LGI}, by performing a detailed comparison of the implementation costs following Ref.~\cite{melegari2025}.

For this purpose, we fix the number of repetitions $N_{\text{shots}}$ of the protocols, and we multiply $N_{\text{shots}}$ by the resources for the single repetition. To determine the three correlators entering the definition of the LG parameter $K = C_{01} + C_{12} - C_{02}$, we have to run the experiment three times. Therefore, to reconstruct $K$, the same procedure has to be repeated $N_{\text{LG}} = 3 N_{\text{shots}}$ times. Instead, concerning non-demolition measurements, each sequence of them is associated with a specific value of $\lambda$. For example, if we take $0 \leq \lambda \leq 100$ with discretization $\delta \lambda = 0.1$, this accounts for a total of $N_{\lambda}=10^3$ experiments. Considering that for each value of $\lambda$, we need to measure both the real and imaginary parts of the quasi-characteristic function $\G$, the total number of protocol repetitions is $N_{\text{QNDM}} = 2 N_{\lambda} N_{\text{shots}}$.

\begin{figure}
    \centering
    \includegraphics[width=0.9\linewidth]{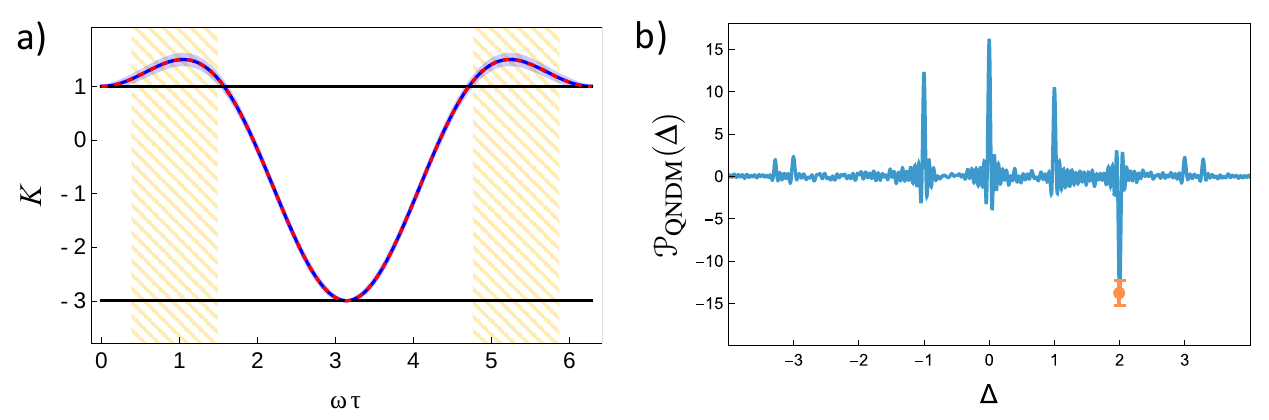}
    \caption{
    a) Simulation results for the computation of the LG parameter $K = C_{01} + C_{12} - C_{02}$ in a scenario where the measurement statistical uncertainty is present, with $N_{\text{shots}} = 10^4$. The solid blue line represents the simulated average curve (obtained over $N_{\text{shots}}$ repetitions), while the dashed red line corresponds to the theoretical prediction as given in the example at the end of Sec.~\ref{sec:QNDM_vs_LGIs}. The blue-shaded region originates from the statistical uncertainty, while the yellow-shaded regions highlight the range of $\omega\tau$ where, considering statistical errors, a violation of the LGI is confidently detected.
    b) Plot of the quasi-probability density functions $\Prob_{\rm QNDM}(\Delta)$ for $\delta\lambda = 1$ and $N_{\text{shots}} = 10^{2}$. These curves are obtained by taking the inverse Fourier transform of the simulated data for $\omega\tau=1.5$, and the orange error bars represent the statistical uncertainty. Notice that, in order to extract single probability values from $\Prob_{\rm QNDM}(\Delta)$, we must integrate it over a fixed interval of $\Delta$. The negative region in $\Prob_{\rm QNDM}(\Delta)$ is visible for any value of $\omega\tau$. These panels are adapted from figures in Ref.~\cite{melegari2025}.
    } 
    \label{fig:QNDM_LGI_comparison_noiseless}
\end{figure} 

Now, let us focus on how statistical uncertainty, due to a finite number of experimental repetitions, affects the accuracy with which a violation of MRps can be identified for both the LG and QNDM protocols. In Fig.~\ref{fig:QNDM_LGI_comparison_noiseless} panel a), we show the LG parameter $K$ obtained by computing the correlators $C_{01}, C_{12}, C_{02}$ in simulation with $N_{\text{shots}} = 10^4$, while in panel b) we plot the QNDM quasi-probability density function $\Prob_{\rm QNDM}(\Delta)$, using $N_{\text{shots}} = 10^{2}$ experimental repetitions with $\delta\lambda = 1$ (thus, $N_{\lambda}=10^2$). This means that, to run the experiments, we need similar resources: $N_{\text{LG}} = 3 \cdot 10^4$ and $N_{\text{QNDM}} = 2 \cdot 10^4$.

In panel a) of Fig.~\ref{fig:QNDM_LGI_comparison_noiseless}, the blue shadowed regions are delineated by the variance associated with the statistical uncertainty of measurements, while the yellow-shaded regions represent the interval of $\omega\tau$ where we are ``statistically certain'' not to satisfy the LGI. The dashed red curve represents the ideal value of $K$ (i.e., with no statistical uncertainty), while the solid blue curve is the mean value of $K$ averaged over the experimental repetitions. In the ideal case, violations of the LGI occur in the ranges $0 \leq \omega \tau \leq \pi/2$ and $3\pi/2 \leq \omega \tau \leq 2\pi$, covering $50\%$ of the total $\omega \tau$ range. The presence of measurement statistical uncertainty significantly reduces the confidence interval for the LGI's violation to $38\%$ of the total $\omega\tau$ range~\cite{melegari2025}.

Panel b) of Fig.~\ref{fig:QNDM_LGI_comparison_noiseless} shows the QNDM quasi-probability density function obtained with $\omega \tau = 1.5$. According to the QNDM protocol, we need to take into account many values of $\G$; in this case, we take $0 \leq \lambda \leq 100$ with discretization $\delta\lambda = 1$. As we can see from the figure, the negative region of $\Prob_{\rm QNDM}(\Delta)$ around $\Delta=2$ is clearly visible, identifying the presence of quantum effects. It turns out that the negative region is visible for any value of $\omega\tau$, albeit in the presence of a measurement statistical uncertainty. With these parameters, the QNDM protocol certifies the violation of MRps for any $\omega\tau$, requiring a comparable amount of resources to those needed to compute the LG parameter $K$. We conclude that, with the same amount of resources, the QNDM always succeeds while the LGI fails in about $60\%$ of the cases.

\begin{figure}
\centering
\includegraphics[width=.8\linewidth]{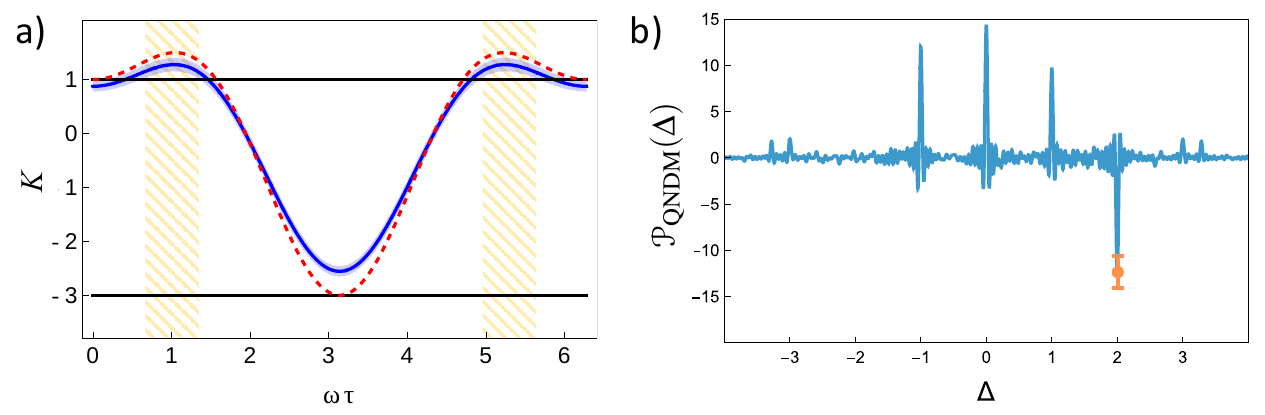}
\caption{
a) Simulation results for the computation of the LG parameter $K = C_{01} + C_{12} - C_{02}$ in the presence of both environmental noise and quantum gate errors for $N_{\text{shots}} = 10^4$.  b) Plot of $\Prob_{\rm QNDM}(\Delta)$ for $\delta\lambda = 1$ and $N_{\text{shots}} = 10^{2}$. In all panels, the notation and parameters are as in Fig.~\ref{fig:QNDM_LGI_comparison_noiseless}. These panels are adapted from figures in Ref.~\cite{melegari2025}.
}
\label{fig:QNDM_LGI_comparison_noise}
\end{figure}

In realistic experiments, we have to take into account also the environmental noise due to the interaction with the external environment, and errors due to the imprecise implementation of the unitary transformations. To see how these noise/error sources can affect the computation of both $K = C_{01} + C_{12} - C_{02}$ and the QNDM quasi-probability density function, we run simulations on a 5-qubit IBM quantum processor using the simulator \texttt{FakeBogotaV2}~\cite{qiskit2024}. The obtained results are presented in Fig.~\ref{fig:QNDM_LGI_comparison_noise}.

Panel a) of Fig.~\ref{fig:QNDM_LGI_comparison_noise} refers to $N_{\text{shots}} = 10^{4}$ repetitions, for a total of $N_{\text{LG}} = 3 \cdot 10^4$ repetitions. From the figure, we can see that the regions where the LGI is not satisfied are further shrunk to $24\%$ of the total $\omega\tau$ range, respectively. On the other hand, panel b) of Fig.~\ref{fig:QNDM_LGI_comparison_noise} displays the QNDM quasi-probability density function obtained with $N_{\text{shots}} = 10^{2}$ and $\delta\lambda = 1$ for a total of $N_{\text{QNDM}} = 2 \cdot 10^4$ repetitions. Despite a quite low level of resources and the presence of environmental noise, the negative region of $\Prob_{\rm QNDM}(\Delta)$ is still clearly visible. Therefore, in this case, the advantage of using QNDM is even more pronounced since, with a fully comparable amount of resources, they always allow for quantumness certification, while the LGI fails roughly in $75\%$ of cases.

\section{Tracking the quantum-to-classical transition}
\label{sec:quantum_to_classical}

A relevant application of the QNDM approach is the tracking of the quantum-to-classical transition due to the interaction with the environment. The quantum-to-classical transition can be identified by looking at the reduction of the negative regions of $\Prob_{\rm QNDM}(\Delta)$~\cite{solinas2021,solinas2022}.

As a prototypical example, let us consider a two-level system $S$ subject to an external drive, and interacting with an external engineered environment $E$. 
The engineered environment allows us to control the coupling strength between the system and the environment. The evolution of the system and the detector is as described above: a sequence of non-demolition measurements at discrete times $t_i$ via a fast, transient coupling with a detector $D$ interspersed with unitary transformations applied to the system's state.

For the sake of clarity, here the physical problem is measuring the work, the internal energy variation, and the heat. Since we are dealing with an open quantum system, the internal energy variation comprises both the work done on the system and the heat dissipated into the environment~\cite{solinas2021,solinas2022}.
To avoid any superfluous complexity, we focus only on the internal energy variation, which is obtained by measuring the system's energy at the start and the end of the thermodynamic transformation~\cite{solinas2021,solinas2022}.

The two-level system evolves under the unitary evolution $\hat{U}_S = \hat{U}_z \hat{U}_x$, where $\hat{U}_z = \exp{(-i \beta \hat{\sigma}_z/\hbar)}$ and $\hat{U}_x = \exp{(-i \alpha \hat{\sigma}_x/\hbar)}$, with $\hat{\sigma}_i$ ($i=x,y,z$) denoting the Pauli operators. This system evolution is generated by the time-dependent Hamiltonian $\hat{H}_S(t)$, where $\hat{H}_S = \epsilon\hat{\sigma}_x/2$ for $t_0 \leq t < t_1$ and $\hat{H}_S = \epsilon \hat{\sigma}_z/2$ for $t_1 < t \leq \Time$. The initial state of the system is $\ket{\psi_0} = \cos(\frac{\theta}{2}) \ket{0}_S + \sin(\frac{\theta}{2}) e^{i \phi}\ket{1}_S$.

The detector $D$ is represented by an additional two-level system whose dynamics is supposed to be frozen during the full evolution of the measured system $S$. The initial state of the detector is $(\ket{0}_D + \ket{1}_D)/\sqrt{2}$.

The Hamiltonian that governs the coupling between the system and the detector is $\hat{H}_{SD}(t) = f(t) \hat{H}_S(t) \otimes \hat{H}_D$, which is operated at times $t_0=0$ and $\Time$. The system-detector coupling at these two times suffices to encode information about the internal energy variation of the system in the phase of the detector state. The time-dependence $f(t)$ in the coupling is chosen in order to implement the unitary transformations $\hat{u}_0 = \exp \{ - (i\lambda/\hbar)~\hat{\sigma}_x \otimes \hat{H}_D\}$ and $\hat{u}_{\Time} = \exp \{ -(i\lambda/\hbar)~\hat{\sigma}_z \otimes \hat{H}_D\}$. We assume that the system-detector coupling occurs over times much smaller than $\epsilon^{-1}$, which is the timescale of the system's dynamics.

In addition, we take into account an engineered open dynamics modelled by an amplitude-damping channel~\cite{nielsen-chuang_book,Garcia-Perez2020}. This means that the environment induces the relaxation of the system from the excited to the ground states with probability $p$, and the relaxation process is associated with a dissipated energy. The parameter $p$ determines the strength of the dissipation process, and it ranges from $p=0$ (no dissipation) to $p=1$ (complete relaxation to the ground state).

To realize the effects of the amplitude-damping channel on the system, we use an additional two-level system, such that the effective system-environment dynamics is described by the transformation
\begin{eqnarray}\label{eq:cold_environment}
\ket{00}_{\rm SE} &\rightarrow& \ket{00}_{\rm SE} \nonumber \\
\ket{10}_{\rm SE} &\rightarrow& \sqrt{1-p} \ket{10}_{\rm SE} + \sqrt{p} \ket{01}_{\rm SE},  
\end{eqnarray}
where the first and the second qubits represent the system and the environment, respectively. 
The transformation in Eq.~(\ref{eq:cold_environment}) models the emission of an energy quantum with probability $p$, whereby we suppose that the environment can only induce relaxation, and not excitation in the system. The relaxation occurs in the basis that diagonalizes $\hat{H}_S(t)$, namely in the basis of $\hat{\sigma}_x$ for $0 \leq t \leq t_1$ and in the basis of $\hat{\sigma}_z$ for $t_1 \leq t \leq \Time$. For simplicity, we denote with $\hat{R}_x$ and $\hat{R}_z$ the operators that implement the transformation (\ref{eq:cold_environment}). Since $\hat{R}_x$ and $\hat{R}_z$ generate two different dissipative dynamics, two distinct engineered environments must be implemented to simulate a Markovian environment: one engineered environment to realize $\hat{R}_x$ and the other for $\hat{R}_z$. Details on how this model can be implemented in a (commercial) superconducting quantum processor can be found in Ref.~\cite{solinas2021}.

\begin{figure}
    \centering
    \includegraphics[scale=.6]{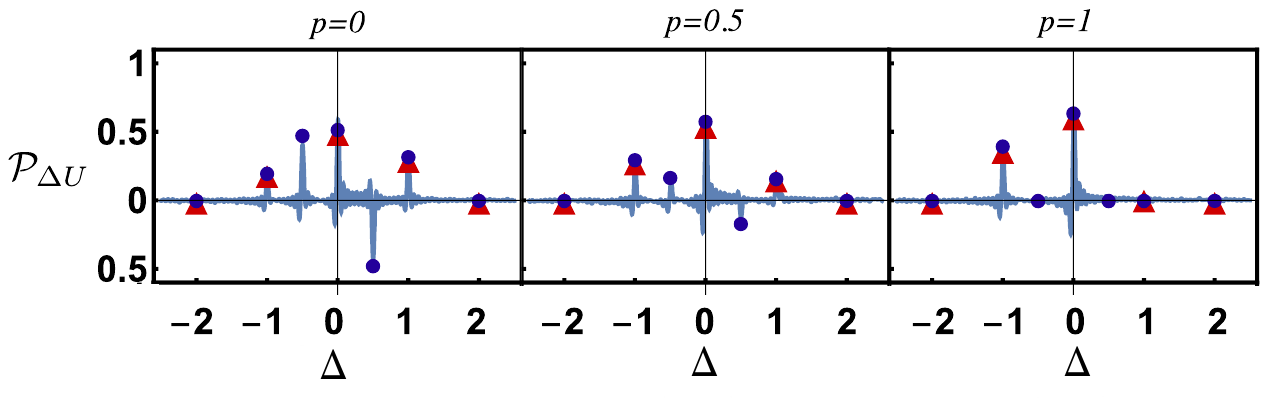}
    \caption{
    QNDM quasi-probability density function $\mathcal{P}_{\Delta U}$ of the internal energy variation $\Delta U$ for a qubit interacting with an environment inducing relaxation. The relaxation strength is parameterized by $p$ that goes from $p=0$ (no dissipation) to $p=1$ (full relaxation process). $\mathcal{P}_{\Delta U}$ is calculated using experimental data obtained from the IBM quantum processor IBMQ-VIGO~\cite{qiskit2024}. The blue dots and the red triangles are the predictions of the QNDM and TPM schemes, respectively. In the experiments, we set $\theta=0.7$, $\phi=1.2$ for the initial state, and $\alpha=1$, $\beta=0.5$ for the system dynamics. The figure is taken from Ref.~\cite{solinas2021}.
    }
    \label{fig:quantum_to_classical_distributions}
\end{figure} 

The total system-detector-environment dynamics is described by the operator $\hat{U}_{\rm diss} = \hat{u}_{\Time}\hat{R}_z\hat{U}_z\hat{R}_x\hat{U}_x\hat{u}_1$. In other terms, first, we couple the system and the detector to measure the energy of the system at the initial time. Then, we let the system evolve with $\hat{U}_x$, and we implement the dissipative dynamics over the $\hat{\sigma}_x$ basis. Afterwards, the last sequence of operations is applied by changing the basis from $x$ to $z$. As we are considering a QNDM scheme, after the whole evolution, we measure the phase accumulated by the detector, and we repeat the procedure by using different couplings $\lambda$ in order to obtain the quasi-characteristic function $\G$. Finally, the inverse Fourier transform is calculated to achieve the quasi-probability density function $\Prob_{\rm QNDM}(\Delta)$.

The results from implementing this procedure in an IBM quantum processor are presented in Fig.~\ref{fig:quantum_to_classical_distributions}.
The solid curve presents the inverse Fourier transform of the experimental data; the blue dots and the red triangles are the QNDM and TPM predictions, respectively. The experimental distribution $\Prob_{\rm QNDM}(\Delta)$ is normalized in order to be compared with the theoretical predictions. As we can see from the figure, the experimental data and theory are in perfect agreement.

From Fig.~\ref{fig:quantum_to_classical_distributions}, it is clear that, when there is no environment, i.e., $p=0$ (panel on the left), $\Prob_{\rm QNDM}(\Delta)$ shows negative regions for the energy exchange of $\epsilon/2$ that is a forbidden transition for the TPM scheme (red dots). When the strength of the environment is increased to $p=0.5$, the negative regions of $\Prob_{\rm QNDM}(\Delta)$ are reduced, as well as the corresponding quantum transition amplitudes; eventually, they vanish when $p=1$. This trend is a signature that the interaction with the environment destroys quantum features of the dynamics, and, thus, it can be interpreted as a quantum-to-classical transition induced by the environment.

\section{Conclusions}
\label{sec:Conclusions}

In this paper, we have reviewed the recent progress in interpreting and utilizing quantum non-demolition measurements to certify the presence of quantum interferences and coherent superpositions in the state of the quantum system under scrutiny.

The QNDM approach has two main advantages with respect to other protocols and tools present in the literature, such as LGIs~\cite{Leggett1985,Clemente2015,ClementePRL2016,HalliwellConference2019}. First, they give a necessary and sufficient condition for the violation of the macrorealism per se, which is directly linked with an operational procedure, discussed in Sec.~\ref{sec:necessary_and_sufficient}. By applying such a procedure, indeed, one is certain to identify --- if present --- quantum coherence in the quantum system's state, independently of the initial state, measurement observables, or time evolution of the system. These violations are certified by examining a single quantity: the negativity of the QNDM quasi-probability density function. Second, QNDM have been successfully implemented in many experimental laboratories worldwide, as one can evince from Refs.~\cite{GeremiaPRL2003,GuerlinNature2007,solinas2021,NiemietzNature2021,solinas2022,YanagimotoPRXQuantum2023}. This is because the QNDM approach only requires coupling the measured system with a detector via a phase operator that has broad experimental applicability. In particular, in Refs.~\cite{solinas2021,solinas2022}, QNDM have been employed to monitor the loss of coherence due to environmental effects in commercial IBM quantum processors. These tests have shown that a QNDM protocol can be realized even in a setting hardly affected by noise. For these reasons, we believe that the approach based on QNDM is a privileged tool to identify quantum features of a system.

To support this claim, in this review paper, we have presented pedagogically key applications of QNDM in different physical scenarios. As the first application, we have surveyed how the QNDM can be exploited to measure the statistics of the work done by a driven quantum system without destroying the quantum interference patterns induced by the work protocol during the system dynamics.
Secondly, we have shown that QNDM can be used as a tool to track the quantum-to-classical transition that a quantum system interacting with an external environment may be forced to undergo.
Finally, we have discussed which is the amount of resources that are needed for the implementation of QNDM. It turns out that the number of experimental repetitions that one has to perform in order to obtain the results predicted by the QNDM approach can be greatly reduced through optimization, without losing significant information about the presence of quantum coherence.

\subsection{What's Next?}

Given the above premises, it is clear that QNDM scheme has numerous applications in various fields of research, ranging from the foundation of quantum mechanics to quantum information and quantumness certifications. However, many studies still need to be undertaken. Below, we list some possible research directions or open points.

\begin{itemize}
    \item     
    Identification of a quantitative relation between the negativity of the QNDM quasi-probability density function and a measure of quantum coherence for the current state of the measured system. The starting point would be Refs.~\cite{solinas2021,solinas2022,solinas2024}. Then, similar ideas might be extended to a measure of entanglement, in relation to local tests of Bell's inequalities~\cite{Calarco1997}.
    \item 
    Detection of the quantum-to-classical transition, using the QNDM approach, for a quantum system approaching mesoscopic and even macroscopic scales. One of the open questions in the foundations of quantum mechanics, indeed, is to understand when a massive quantum body loses its quantum features, thus starting to behave classically. This line of research is still under investigation, as proved by Refs.~\cite{Arndt1999,Gerlich2011,Aspelmeyer2022,FuchsScienceAdvances2024,Pedalino2026}. In particular, in recent years, double-slit experiments have been performed using different bodies with an increasing mass. 
    Since the QNDM tracks quantum interferences and coherent superpositions, it is worth testing experimentally possible violations of macrorealism {\it per se}, according to QNDM quasi-probabilities, in levitated mechanical systems of mesoscopic size, as in Ref.~\cite{FuchsScienceAdvances2024}.
    \item 
    The QNDM protocol resembles the decoherent histories approach to quantum theory, introduced by Gell-Mann, Hartle, Griffiths, Omnes in the mid-nineties~\cite{Griffiths1984,Gell-Mann1990,Gell-Mann1993,Griffiths1993,Omnes1999,Gell-Mann2012,Gell-Mann2014} and discussed recently in Refs.~\cite{Hohenberg2010,Strasberg2024,Wang2025}. Indeed, the decoherence functional in these cited papers comprises exactly the interference contributions in Eq.~\eqref{eq:ProbAmplitude} for the quantum system to evolve along superposed trajectories. Moreover, the classical behaviour of the system under scrutiny obtained when the QNDM quasi-probability density function is positive [see Eq.~\eqref{eq:Pcl_Delta}] corresponds to the decoherence functional going to zero. In this perspective, the QNDM protocol could be used to experimentally test the results and predictions of the decoherent histories approach. However, we must stress that the number of possible quantum trajectories to be tracked during the dynamics increases rapidly with the number of measurements performed. This limits the applicability of QNDM, because the quasi-probability spectrum soon becomes too complex to disentangle the contributions of the different trajectories.
    \item 
    Application of QNDM to quantum computational complexity. In the field of quantum computing, one of the most important open questions is which fundamental resources enable a quantum advantage to be achieved. Several studies suggest that entanglement~\cite{Horodecki2009}, nonstabilizerness~\cite{Leone2022}, and quantum coherence~\cite{Baumgratz2014,Streltsov2017} are all fundamental. Recently, Thomas {\it et al.}~\cite{Thomas2025} have shown that the ``path coherence'', i.e., the interference of different paths developed during a quantum computation, is a fundamental measure to estimate the hardness of the computation and, thus, the impossibility to perform such a calculation with a classical computer. The measurement to achieve such ``path coherence'' might be directly related to the quasi-probability density function that arises from the QNDM protocol. While the results in Ref.~\cite{Thomas2025} are exclusively theoretical, QNDM could be used to experimentally measure ``path coherence'' in quantum computers.
    \item
    The most used quasi-probability so far is the Wigner function (WF)~\cite{Hillery1984}. As for QNDM quasi-probabilities, negative regions of the WF are used as a witness of quantum phenomena. Thus, it will be important to explore the connection between these two families of quasi-probabilities. Such a link could lead to new insights about the negativity of the WF.
    \item 
    Computation of quantum speed limit bound for non-demolition quasi-probabilities, taking inspiration from Ref.~\cite{PratapsiQST2025} that addresses this issue for Kirkwood-Dirac quasi-probabilities.
    \item 
    Extension of the proof of the sufficient and necessary condition for the violation of the macrorealism per se to open quantum systems. In this regard, by setting ourselves in a context of monitoring heat generation, one could certify the presence of quantum contributions in the heat distribution using QNDM quasi-probabilities.     
\end{itemize}

\section*{Acknowledgments}

S.G.~acknowledges financial support from the PNRR MUR project PE0000023-NQSTI funded by the European Union---Next Generation EU, and from the PRIN project 2022FEXLYB Quantum Reservoir Computing (QuReCo).

\bibliographystyle{apsrev}
\bibliography{QuasiProbabilities}

\end{document}